\documentclass{LPEP}

\pdfoutput=1

\newcommand{\bfmu} {\boldsymbol{\mu}}

\newcommand{\bftheta} {\boldsymbol{\theta}}
\newcommand{\bfomega} {\boldsymbol{\omega}}

\newcommand{\bfSigma} {\boldsymbol{\Sigma}}

\newcommand{\bfOmega} {\boldsymbol{\Omega}}

\newcommand{\bzero}{ {\boldsymbol 0} }

\newcommand{\bb}{ {\boldsymbol b} }

\newcommand{\bd}{ {\boldsymbol d} }

\newcommand{\bE}{ {\boldsymbol E} }

\newcommand{\bH}{ {\boldsymbol H} }

\newcommand{\bJ}{ {\boldsymbol J} }

\newcommand{\bm}{ {\boldsymbol m} }

\newcommand{\bv}{ {\boldsymbol v} }
\newcommand{\bV}{ {\boldsymbol V} }

\newcommand{\bW}{ {\boldsymbol W} }
\newcommand{\bx}{ {\boldsymbol x} }
\newcommand{\bX}{ {\boldsymbol X} }
\newcommand{\by}{ {\boldsymbol y} }

\newcommand{\bz}{ {\boldsymbol z} }

\newcommand{\bfy} {\mathbf{y}}

\renewcommand{\Pr}{\mathsf{Pr}}

\newcommand{\reals}{\mathbb{R}}

\DeclareMathOperator{\diag}{diag}

\newcommand{\normal}{\mathsf{N}}

\newcommand{\bbeta}{\boldsymbol{\beta}}
\newcommand{\bld}[1]{\boldsymbol{#1}}
\newcommand{\bgam}{\boldsymbol{\gamma}}
\newcommand{\betag}{\boldsymbol{\beta}_{\boldsymbol{\gamma}}}

\usepackage{multirow}

\setcitestyle{round}
\DeclareMathOperator*{\argmax}{arg\,max}

\begin{document}

\begin{center}
  {\LARGE Laplace Power-expected-posterior priors for generalized linear models with applications to logistic regression}\\\ \\
  {Anupreet Porwal and Abel Rodriguez\\ 
    Department of Statistics, University of Washington Seattle, WA, 98195, USA
  }
\end{center}

\begin{abstract}
  Power-expected-posterior (PEP) methodology, which borrows ideas from the literature on power priors, expected-posterior priors and unit information priors, provides a systematic way to construct objective priors. The basic idea is to use imaginary training samples to update a noninformative prior into a minimally-informative prior. In this work, we develop a novel definition of PEP priors for generalized linear models that relies on a Laplace expansion of the likelihood of the imaginary training sample. This approach has various computational, practical and theoretical advantages over previous proposals for non-informative priors for generalized linear models. We place a special emphasis on logistic regression models, where sample separation presents particular challenges to alternative methodologies.  We investigate both asymptotic and finite-sample properties of the procedures, showing that is both asymptotic and intrinsic consistent, and that its performance is at least competitive and, in some settings, superior to that of alternative approaches in the literature.
\end{abstract}

\section{Introduction}

Generalized linear models (GLMs, e.g., see \citealp{mccullagh2019generalized}) are one of the main workhorses of statistical analysis. Indeed, GLMs are widely used both to model data directly and as building blocks for more complex hierarchical models.  However, in spite of their broad adoption, prior elicitation for general GLMs in the absence of subjective information remains an open problem, particularly in settings where the main goal is variable selection.  Because standard non-informative priors for GLMs that work well for parameter estimation are often improper, they cannot be used in model selection problems because they typically lead to ill-defined Bayes factors (e.g., see \citealp{berger2001objective}).

Within the subclass of Gaussian linear models, the literature on so-called ``objective'' or ``default'' priors is extensive.  Examples include point-mass spike-\&-slab priors \citep{mitchell1988bayesian,geweke1996variable}, $g$-priors \citep{Zellner1986}, mixtures of $g$-priors \citep{zellner1980posterior,liang2008mixtures}, unit information priors \citep{kass1995reference}, intrinsic Bayes factors \citep{berger1996linear}, fractional Bayes factors \citep{o1995fractional,de2001consistent}, non-local priors \citep{johnson2010use,johnson2012bayesian} and power-expected-posterior priors \citep{fouskakis2015power}, among other approaches.  See \cite{bayarri2012criteria} for a review and a discussion of desirable properties.  
The literature on default priors for GLMs is more limited, with three main approaches dominating. These include those introduced by \cite{bove2011hyper} and  \cite{li2018mixtures}, both of which consider modifications of mixtures of g-priors that are suitable for GLMs, and \cite{fouskakis2018power}, who considers extensions of power-expected-posterior priors that relies on unnormalized power likelihoods.  One feature shared by all three approaches is that they can be thought of as being based on the idea of calibrating (possibly improper) priors using either real or imaginary training samples (e.g., see \citealp{berger1996intrinsic} and \citealp{perez2002expected}).

In this paper we introduce a variant of the power-expected-posterior (PEP) prior for GLMs that we call the Laplace PEP, or LPEP.  While the formulation is general, this manuscript emphasizes the development of the LPEP for logistic regression models.  This is because this subclass of models provides the best illustration of the theoretical and practical advantages of our approach over existing ones.  For example, we note that the prior described in \cite{li2018mixtures} is improper when the maximum likelihood estimator of the regression coefficients under the observed data does not exist.  In the case of logistic regression, this happens when there is separation among the groups (e.g., see \citealp{albert1984existence}, \citealp{lesaffre1989partial} and \citealp{heinze2002solution}).  Separation is reasonably common in practical applications, especially in problems with relatively small samples and several unbalanced and highly predictive risk factors (e.g., see Section \ref{se:urinary}).  A similar issue arises with the PEP priors introduced in \cite{fouskakis2018power} since the imaginary training samples are not restricted to yield finite maximum likelihood estimators.  Furthermore, both versions of the PEP prior proposed by \cite{fouskakis2018power} are computationally intractable, requiring the use of reversible Jump Markov Chain Monte Carlo algorithms \citep{green1995reversible,dellaportas2002bayesian}. Our LPEP is carefully designed to deal with separation issues, and it is well defined as long as at least one training sample exists that yields finite maximum likelihood estimates under the full model.  Furthermore, the LPEP can be written as a location-and-scale mixture of Gaussian priors and is therefore computationally tractable and easy to incorporate into standard Markov chain Monte Carlo algorithms that rely on data augmentation (e.g., \citealp{polson2013bayesian}). Finally, the mixture structure also simplifies the theoretical study of the prior, allowing us to show that the procedure is both asymptotically consistent and intrinsically consistent.

It is important to stress that the focus of this manuscript is on priors for variable selection that place positive probability on specific coefficients being exactly zero.  An alternative approach is to use continuous shrinkage priors.  Examples include the Bayesian Lasso \citep{park2008bayesian}, the Horseshoe prior \citep{carvalho2010horseshoe}, the Normal-Gamma prior \citep{brown2010inference}, the Dirichlet-Laplace prior \citep{bhattacharya2015dirichlet}, global-local shrinkage priors \citep{polson2012local}, the Beta-prime prior \citep{bai2018beta}, the tail-adaptive shrinkage prior \citep{lee2020continuous} and the Horseshoe-pit prior \citep{denti2021horseshoe}.  Continuous shrinkage priors tend to have computational advantages and are very effective in predictive settings.  However, because they place  probability zero on any one value of the parameter space, variable selection can be performed only by thresholding the posterior distributions of the model coefficients.  While ad-hoc techniques have been devised for this purpose (e.g., see \citealp{li2017variable}), thresholding tends to work well mostly in settings where enough prior information is available to establish practical significance.

The remainder of the paper is organized as follows:  Section \ref{sec:PEPreview} reviews the general definition and properties of PEP priors.  Section \ref{sec:LPEP} defines the general LPEP and conditions under which it is well defined.  Section  \ref{se:LPEPlogistic} discusses in detail the special case of the LPEP for logistic regression, including  theoretical properties such as intrinsic and model selection consistency and computational strategies.  Sections \ref{se:simulation} and \ref{se:realdata} presents empirical results from simulation studies as well as three different real datasets.  Finally, Section \ref{se:discussion} discusses our results and future directions for research.

\section{Power-expected-posterior priors:  A brief review}\label{sec:PEPreview}

Power-expected-posterior (PEP) priors \citep{fouskakis2015power} extend the expected-posterior (EP) priors introduced by \cite{perez2002expected} by controlling the amount of information contained in the prior using the prior power approach originally developed by  \cite{ibrahim2000power} and \cite{chen2000power} in the context of subjective priors based on historical information.

Briefly, let $\by$ denote the $n$-dimensional vector containing the observed data, $\bgam$ index the model space, and $\bbeta_{\bgam}$ represent vector of parameters under model $\bgam$.  We start with a (potentially improper) prior $\pi_{\bgam}^N(\bbeta_{\bgam})$ under model $\bgam$ and introduce an $n^{*}$-dimensional vector of \textit{imaginary} training samples arising from a distribution $m^{*}(\bfy^{*})$.  The EP prior is then constructed as
\begin{align*}
    \pi_{\bgam}^{EP}(\bbeta_{\bgam})=\int
    \frac{f_{\bgam}\left(\by^{*} \mid \bbeta_{\bgam}\right) \pi_{\bgam}^N\left(\bbeta_{\bgam}\right)}{\int  f_{\bgam}\left(\by^{*} \mid \bbeta_{\bgam}\right) \pi_{\bgam}^N\left(\bbeta_{\bgam}\right) d \bbeta_{\bgam}}
    m^{*}(\by^{*}) d\by^{*}
\end{align*}
In words, the EP priors use the imaginary training sample $\bfy^{*}$ to update the original prior $\pi_{\bgam}^N(\bbeta_{\bgam})$, and addresses the possible effect of using any particular training sample by averaging over the distribution $m^{*}(\bfy^{*})$.  The use of a common $m^{*}(\by^{*})$ properly calibrates the priors across the different values of $\bgam$, even in situations where $m^{*}(\by^{*})$ is improper. \cite{perez2002expected} discuss various possible choices of $m^{*}(\by^{*})$ in both informative and non-informative settings.

Note that an implicit assumption in the formulation of the PEP is that the training sample $\bfy^{*}$ must be large enough so that the posterior based on it is proper, i.e., 
\begin{align}\label{eq:proper}
\int  f_{\bgam}\left(\by^{*} \mid \bbeta_{\bgam}\right) \pi_{\bgam}^N\left(\bbeta_{\bgam}\right) d \bbeta_{\bgam} < \infty
\end{align}
for any $\by^{*}$ in the support of $m^{*}$.  However, large the values of $n^{*}$ will produce priors that are relatively concentrated.  To balance these two goals, it is common to choose $n^{*}$ as the size of the minimum training sample required to satisfy \eqref{eq:proper} across all models.  This complicates the implementation of this prior in situations where data is not independent and identically distributed, such as in regression models. 

In summary, even though the EP prior attempts to ameliorate the effect of the $\by^{*}$ by averaging over $m^{*}$ and by using training samples that are as small as possible, in some applications the prior might be quite concentrated, and therefore highly informative.  Power-expected-posterior priors \citep{fouskakis2015power} address this drawback by scaling the  likelihood of the imaginary sample,
\begin{align*}
    \pi_{\bgam}^{PEP}(\bbeta_{\bgam})=\int
    \frac{\tilde{f}_{\bgam}\left(\by^{*} \mid \bbeta_{\bgam}, {\delta}\right) \pi_{\bgam}^N\left(\bbeta_{\bgam}\right)}{\int  \tilde{f}_{\bgam}\left(\by^{*} \mid \bbeta_{\bgam}, {\delta}\right) \pi_{\bgam}^N\left(\bbeta_{\bgam}\right) d \bbeta_{\bgam}}
    m^{*}(\by^{*} \mid \delta) f(\delta \mid \bgam) d\delta d\by^{*}
\end{align*}
where $\tilde{f}_{\bgam}\left(\by^{*} \mid \bbeta_{\bgam}, {\delta}\right) = \frac{f_{\bgam}\left(\by^{*} \mid \bbeta_{\bgam}\right)^{\frac{1}{\delta}}}{\int f_{\bgam}\left(\by^{*} \mid \bbeta_{\bgam}\right)^{\frac{1}{\delta}}d\bbeta_{\bgam}}$ is the normalized power likelihood for the training sample $\by^{*}$ based on model $\bgam$, and $\delta$ is the power parameter.  If $\delta=1$, then PEP prior reduces to the EP prior, while values of $\delta > 1$ yield priors with a larger variance (and therefore, less information) than the EP prior.  A particularly appealing choice is $\delta = n^{*}$ (or, alternatively, a prior on $\delta$ that is concentrated around $n^*$), which leads to a prior that can be considered as being unit information \citep{kass1995reference}.  Note that $\delta$ plays a similar role to the $g$ parameter involved in the definition of (mixtures of) $g$ priors, so that treating $\delta$ as random will typically lead to priors that have heavier tails, and are therefore more robust (in the sense of \citealp{dawid1973posterior} and \citealp{andrade2011bayesian}).

Being able to use the parameter $\delta$ to control the amount of information contained in the prior means that the choice of the size of training sample is less critical in the case of PEP priors.  In the sequel, we work with $n^{*}=n$, a choice that is particularly convenient when dealing with GLMs and other regression models.  Indeed, taking $n^{*}=n$ allows us to select $\bX^{*}$, the design matrix associated with the training sample $\by^{*}$, as $\bX^{*} = \bX$,  the design matrix associated with the observed data.  A further implicit assumption moving forward is that $n > p$.

The PEP prior was originally derived for model selection in Gaussian linear model.  In that case, 
computing the normalizing constant $\int f_{\bgam}\left(\by^{*} \mid \bbeta_{\bgam}\right)^{\frac{1}{\delta}}d\bbeta_{\bgam}$ associated with $\tilde{f}_{\bgam}\left(\by^{*} \mid \bbeta_{\bgam}, {\delta}\right)$ is straightforward.  Indeed, for most standard choices of $\pi_{\bgam}^N\left(\bbeta_{\bgam}\right)$, the induced PEP can be written as a location-and-scale mixture of Gaussian distributions, dramatically simplifying computation within a Markov chain Monte Carlo framework.  This property, however, does not extend to other GLMs.  To address this issue, \cite{fouskakis2018power} introduce two slightly different modifications of the PEP framework that rely on the unnormalized power likelihood $f_{\bgam}\left(\by^{*} \mid \bbeta_{\bgam}\right)^{\frac{1}{\delta}}$ rather than $\tilde{f}_{\bgam}\left(\by^{*} \mid \bbeta_{\bgam}, {\delta}\right)$: the concentrated reference PEP (CRPEP) and the diffuse reference PEP (DRPEP).  However, while the use of the unnormalized power likelihood avoids some of the computational difficulties associated with the original PEP prior, many of them remain.  In particular, neither $\pi_{\bgam}^{CRPEP}(\bbeta_{\bgam})$ nor $\pi_{\bgam}^{DRPEP}(\bbeta_{\bgam})$ belong to standard families of distributions.  This prevents closed-form integration of the regression coefficients and therefore requires the use Reversible Jump Markov chain Monte Carlo algorithms.  Furthermore, the definition of the CRPEP and the DRPEP and the computational approach introduced by the authors (which relies on Laplace approximations to compute certain normalizing constants needed for the acceptance probabilities of various Metropolis-Hastings steps) implicitly assume that the maximum likelihood estimate of $\bbeta_{\bgam}$ exists for any training sample $\by^{*}$ and model $\bgam$.  However, this constraint, which as we discussed in the introduction might be binding for some classes of GLMs, is not accounted for in the definition of the $m^{*}(\by^{*} \mid \delta)$ CRPEP and DRPEP.


\section{The Laplace power-expected-posterior prior for Generalized Linear Models}\label{sec:LPEP}

Instead of working with the unnormalized power likelihood as in \cite{fouskakis2018power}, in this paper we propose replacing the likelihood of the imaginary samples with its Laplace approximation \textit{before} raising it to the power $1/\delta$.  Hence, the name Laplace PEP, or LPEP.  More concretely, let the observations $\by =(y_1, \ldots, y_n)^T$ be generated from a likelihood of the form
$$
f_{\bgam}\left(\by \mid \bbeta_{\bgam}\right) =  \prod_{i=1}^{n} h(y_i, \tau) \exp\left\{ \frac{T(y_i) G\left(\eta\left(\bx_{\bgam,i}'\bbeta_{\gamma}\right)\right) + A\left(\eta\left(\bx_{\bgam,i}'\bbeta_{\gamma}\right)\right)}{d(\tau)} \right\}
$$
where $E(y_i) = \eta(\bx_{\bgam,i}'\bbeta_{\gamma}) = \theta_i$ for some appropriate link function $\eta$, $\bx_{i}=(1,x_{i,1},\dots,x_{i,p})$ is the $p+1$ dimensional vector of regressors associated with observation $y_i$, $\bbeta \in S$ is the $p+1$ dimensional vector of regression coefficients (including the intercept), $S$ is a connected open subset of $\reals^{p+1}$, $\bgam' = (\gamma_0, \gamma_1, \ldots, \gamma_p)$ is a binary vector of length $p+1$ such that for all $j\in\{1,\dots,p\}, \gamma_j=1$ if the $j$-th variable is included in the model (i.e., if $\beta_j$ is different from zero) and $\gamma_j=0$ otherwise and $\gamma_0 =1$ (i.e., intercept is always included in the model),  $\bx_{\bgam,i}$ and $\bbeta_{\gamma}$ denote the sub-vectors of $\bx_{i}$ and $\bbeta$ with length $p_{\bgam}+1$ where $p_{\bgam} = \sum_{j=1}^{p} \gamma_j$ that include only those components for which the corresponding $\gamma_j$ is equal to 1, and $\tau$ is an overdispersion parameter.  In order to simplify our exposition, and following standard practice in the literature, in the sequel we treat $\tau$ as known.  The generalization of our approach to situations where $\tau$ is unknown is relatively straightforward as $\tau$ will typically be a parameter that is common to all models under consideration, and can therefore be safely assigned a standard (potentially improper) non-informative prior \citep{berger1998bayes}.  When $\tau$ is known, the normalizing constant $h(y_i, \tau)$ can be dropped, and the scaling function $d(\tau)$ can be absorbed within the function $T$ and $A$, leading to the somewhat simpler expression:
\begin{align}\label{eq:likelihoodGLM}
f_{\bgam}\left(\by \mid \bbeta_{\bgam}\right) &=   \exp\left\{ \sum_{i=1}^{n} \left[ T^{*}(y_i) G\left(\eta\left(\bx_{\bgam,i}'\bbeta_{\gamma}\right)\right) + A^{*}\left(\eta\left(\bx_{\bgam,i}'\bbeta_{\gamma}\right)\right) \right] \right\} ,
\end{align}
where $A^{*}(\cdot) = A(\cdot)/d(\tau)$ and $T^{*}(\cdot)=T(\cdot)/d(\tau)$.

The second order Laplace approximation to \eqref{eq:likelihoodGLM} is given
by
\begin{align}\label{eq:LaplacelikelihoodGLM}
f_{\bgam}\left(\by \mid \bbeta_{\bgam}\right) & \approx f^{L}_{\bgam}\left(\by \mid \bbeta_{\bgam}\right) \propto  \exp\left\{ - \frac{1}{2} \left(\bbeta_{\gamma} - \hat{\bbeta}_{\gamma}(\by)\right)' \bH_{\bgam}(\by) \left(\bbeta_{\gamma} - \hat{\bbeta}_{\gamma}(\by)\right) \right\} ,
\end{align}
where $\tilde{\ell}(\bbeta_{\bgam}) = \sum_{i=1}^{n} \left[ T^{*}(y_i) G\left(\eta\left(\bx_{\bgam,i}'\bbeta_{\gamma}\right)\right) + A^{*}\left(\eta\left(\bx_{\bgam,i}'\bbeta_{\gamma}\right)\right) \right]$, $\hat{\bbeta}_{\bgam}(\by)$ denotes the maximum likelihood estimate for $\bbeta_{\bgam}$ based on sample $\by$, and $\bH_{\bgam}(\by)$ is the $(p_{\bgam} +1) \times (p_{\bgam}+1)$ observed information matrix with entries
$$
[\bH_{\bgam}(\by)]_{j,j'} = - \left. \frac{\partial }{\partial \beta_{\bgam,j} \partial \beta_{\bgam,j'}} \tilde{\ell}(\bbeta_{\bgam}) \right|_{\bbeta_{\bgam} = \hat{\bbeta}_{\bgam}(\bfy)} .
$$
In the case of regular exponential families, it is well known that this approximation is accurate up to an $\mathcal{O}(\frac{1}{n})$ order term (e.g., see \citealp{schwarz1978estimating} and \citealp{haughton1988choice}).  With this in mind, we define the LPEP as
\begin{align}\label{eq:PEPGLM}
    \pi_{\bgam}^{LPEP}(\bbeta_{\bgam})=\int
    \frac{\tilde{f}^{L}_{\bgam}\left(\by^{*} \mid \bbeta_{\bgam}, {\delta}\right) \pi_{\bgam}^N\left(\bbeta_{\bgam}\right)}{\int  \tilde{f}^{L}_{\bgam}\left(\by^{*} \mid \bbeta_{\bgam}, {\delta}\right) \pi_{\bgam}^N\left(\bbeta_{\bgam}\right) d \bbeta_{\bgam}}
    m^{*}(\by^{*} \mid \bX) f(\delta \mid \bgam) d\delta d\by^{*} ,
\end{align}
where $\bX$ is the $n \times (p+1)$ design matrix whose rows correspond to the $\bx_{i}'$ vectors.

Defining the LPEP using \eqref{eq:LaplacelikelihoodGLM} instead of \eqref{eq:likelihoodGLM} dramatically simplifies computation.  Indeed, using \eqref{eq:LaplacelikelihoodGLM} implies that $\tilde{f}^{L}_{\bgam}\left(\by^{*} \mid \bbeta_{\bgam}, {\delta}\right)$ is proportional to a Gaussian kernel,
$$
\tilde{f}_{\bgam}\left(\by^{*} \mid \bbeta_{\bgam}, {\delta}\right) \propto \delta^{-\frac{p_{\gamma}+1}{2}}
\exp\left\{ - \frac{1}{2\delta} \left(\bbeta_{\gamma} - \hat{\bbeta}_{\gamma}(\by^{*})\right)' \bH_{\bgam}(\by^{*}) \left(\bbeta_{\gamma} - \hat{\bbeta}_{\gamma}(\by^{*})\right) \right\} ,
$$
and, therefore, for standard choices of $\pi_{\bgam}^N\left(\bbeta_{\bgam}\right)$ (such as the flat prior $\pi_{\bgam}^N\left(\bbeta_{\bgam}\right) \propto 1$), $\pi^{LPEP}$ corresponds to a location-and-scale mixture of Gaussian distributions.

Besides using the the Laplace approximation of $f(\by \mid \bbeta_{\bgam})$, the definition in \eqref{eq:PEPGLM} differs from that in \cite{fouskakis2018power} in terms of the structure of the distribution of the imaginary samples $m^{*}$ in two important ways.  First, note that we \textit{do not} make $m^{*}$ dependent of the scaling factor $\delta$.  This makes intuitive sense (there is no obvious reason why the power factor used to re-scale the information in the training sample should also affect how the training sample is generated) and simplifies both posterior computation and theoretical analysis.  On the other hand, $m^{*}$ is made to depend explicitly on the design matrix $\bX$.  In particular, we take 
\begin{align}\label{eq:mtraining}
m^{*}(\by^{*} \mid \bX) &\propto \tilde{m}^{*}(\by^{*} \mid \bX) \mathbf{1}\left(\bfy^{*} \in A(\bX)\right) ,
\end{align}
where $\mathbf{1}\left(\bfy^{*} \in A(\bX)\right)$ is the indicator function on the set $A(\bX)$ and
\begin{align}\label{eq:constraint}
A(\bX) &= \left\{ \tilde{\by} \mid \hat{\bbeta}_{\bgam}\left(\tilde{\by}\right) \mbox{ exists and is finite for all } \bgam 
\right\} .
\end{align}

At first sight, the computational implementation of \eqref{eq:constraint} might seem daunting, as it in principle requires that the existence of the maximum likelihood estimator of the parameters be checked for every possible model under consideration.  However, as the following theorem shows, for a broad class of GLMs that includes binomial and multinomial models with logistic and probit links, as well as Poisson models with the canonical logarithmic link, it is enough to check relatively simple conditions on the loglikelihood funciton for the full model.

\begin{theorem}\label{th:existence}
Let $\ell_{\bgam}(\bbeta_{\bgam})$ denote the log-likelihood function of the GLM in \eqref{eq:likelihoodGLM} associated with model $\bgam$ and $\bgam_F = (1,1, \ldots,1)$ denote the full model (i.e., the model that includes all potential regressors).
Assume that 
\begin{enumerate}
    \item[(i)] $\ell_{\bgam_F}(\bbeta_{\bgam_F})$ is continuous and striclty concave on $S_{\bgam_F}$
    \item[(ii)] $\lim_{\bbeta_{\bgam_F} \to \partial S_{\bgam_F}}\ell_{\bgam_F}(\bbeta_{\bgam_F}) = -\infty$, where $\partial S_{\bgam_F}$ represents the closure of $S_{\bgam_F}$
\end{enumerate}
Then, $\hat{\bbeta}_{\bgam}(\by)$, the maximum likelihood estimator, exists under any other model $\bgam$.
\end{theorem}

The proof of Theorem \ref{th:existence}, which is relatively straightforward, can be seen in Appendix \ref{ap:MLEinsubmodels}.  Similar theorems covering wider classes of GLMs can be derived using the results in \cite{wedderburn1976existence}, but this version suffices for our purposes.

To conclude, it might be helpful to expand on the relationship between the LPEP prior as we defined it in this Section and other similar proposals in the literature within the context of Gaussian linear models.  Note that, while the Laplace approximation in \eqref{eq:LaplacelikelihoodGLM} is exact for the Gaussian linear model, our definition of the LPEP is not equivalent to the PEP prior in \cite{fouskakis2015power} because the predictive distribution $m^{*}(\by^{*})$ used the generate the training samples does not depend on $\delta$.  On the other hand, the LPEP for the Gaussian linear model looks similar to a mixture of $g$ priors \cite{liang2008mixtures}.  In particular, while the covariance matrix of the Gaussian kernels involved is given by $\delta \left\{ \bX^T\bX\right\}^{-1}$ in both priors, the conditional mean of the LPEP depends on the training sample and is not zero.  However, depending on the choice of $m^{*}(\bfy
^{*})$, the intrinsic prior associated with the LPEP does correspond to a mixture of $g$ priors (please see Section \ref{se:intrinsic}).

\section{The LPEP prior for logistic regression}\label{se:LPEPlogistic}

In the sequel, we illustrate the LPEP in the context of logistic regression, where the likelihood function for model $\bgam$ can be written as
\begin{align}\label{eq:liklogistic}
     f_{\bgam}(\bld{y} \mid \betag)= \exp \left\{ \sum_{i=1}^n \left[ \bld{x}_{\bgam,i}^T\betag - \log (1 + \exp\left\{ \bld{x}_{\bgam,i}^T\betag \right\}) \right]
     \right\} ,
\end{align}
and $S_{\bgam} = \reals^{p_{\bgam}+1}$.  A natural choice for $\pi^N_{\bgam}(\bbeta_{\bgam})$ in this setting is the (improper) flat prior $\pi^N_{\bgam}(\bbeta_{\bgam}) \propto 1$, so that 
%
\begin{align}\label{eq:LPEPlogistic}
\pi^{LPEP}_{\bgam}(\bbeta_{\bgam}) = \sum_{\by^* \in \{0,1\}^n} \Bigg[ \int \phi_{p_{\bgam}+1} 
 \left( \bbeta_{\bgam} \mid  \hat{\bbeta}_{\bgam}(\by^{*}) , \delta \bH^{-1}_{\bgam}\left(\by^{*}\right) \right)
 f(\delta \mid \bgam) d\delta \Bigg] m^{*}(\by^{*} \mid \bX) ,
\end{align}
where $\phi_{p}\left(\cdot \mid \bfmu, \bfSigma\right)$ denotes the density of the $p$-variate normal distribution with mean vector $\bfmu$ and covariance matrix $\bfSigma$, $\hat{\bbeta}_{\bgam}(\by^{*})$ is the maximum likelihood estimator of $\bbeta_{\bgam}$ based on the training sample $\by^{*}$ (which, while not available in closed form, can be easily evaluated),
$$
\bH_{\bgam}(\by^{*}) = \sum_{i=1}^{n} \frac{1}{\hat{\theta}_{\bgam,i}(\by^{*}) (1-\hat{\theta}_{\bgam,i}(\by^{*}))} \bx_{\bgam,i} \bx_{\bgam,i}' = \bX_{\bgam}' \bW_{\bgam}(\by^{*}) \bX_{\bgam} ,
$$
where
$$
\hat{\theta}_{\bgam,i}(\by^{*}) = \frac{1}{1 +  \exp\left\{ \bld{x}_{\bgam,i}^T\hat{\bbeta}_{\bgam}(\by^{*}) \right\}} ,
$$
and
$$\bW_{\bgam}(\by^{*}) = \diag \left\{ \frac{1}{\hat{\theta}_{\bgam,1}(\by^{*}) (1-\hat{\theta}_{\bgam,1}(\by^{*}))}, \ldots, \frac{1}{\hat{\theta}_{\bgam,n}(\by^{*}) (1-\hat{\theta}_{\bgam,n}(\by^{*}))} \right\}.
$$

Note that the loglikelihood function $\ell_{\bgam}(\bbeta_{\bgam})=\sum_{i=1}^n \left[ \bld{x}_{\bgam,i}^T\betag - \log \left(1 + \exp\left\{ \bld{x}_{\bgam,i}^T\betag \right\}\right) \right]$ associated with \eqref{eq:liklogistic} is continuous everywhere for any model $\bgam$ (and, in particular, for $\bgam_{F}$).  Furthermore, as long as the full design matrix $\bX$ is full rank, $\bH_{\bgam}(\by^{*})$ is strictly positive definite for any $\bgam$.  Hence, the model satisfies condition (i) in
Theorem \ref{th:existence}.  To verify that condition (ii) is satisfied, it is enough to show that the training sample $\by^{*}$ is not separable under the full design matrix $\bX$.  \cite{Konis2007linear} discusses an efficient approach to detect separation in logistic regression models that relies on linear programming.  This approach has been implemented in the \texttt{R} package \texttt{detectseparation} \citep{Kosmidis2020detect}.

We discuss next the choice of $\tilde{m}^*(\by^{*})$ in \eqref{eq:mtraining}.  A common approach is to select $m^{*}(\by^{*})$ as the predictive under the the null model. In that spirit, for the logistic regression model (and, more generally, for any binary regression model, independently of the link function used) we set 
\begin{align*}
\tilde{m}^*(\by^{*}) &= \frac{\Gamma\left( \sum_{i=1}^{n}y_{i}^{*} +  \frac{1}{2}\right)\Gamma\left(n - \sum_{i=1}^{n}y_{i}^{*} + \frac{1}{2}\right)}{\Gamma(n + 1)\Gamma\left( \frac{1}{2}\right)\Gamma\left( \frac{1}{2}\right)} \\
 & = \int t^{\sum_{i=1}^{n}y_{i}^{*}}  (1 - t)^{n - \sum_{i=1}^{n}y_{i}^{*}}  \, \frac{t^{-\frac{1}{2}} (1-t)^{-\frac{1}{2}}}{\Gamma\left( \frac{1}{2}\right)\Gamma\left( \frac{1}{2}\right)} d t,
\end{align*}
a Beta-Binomial distribution with both parameters equal to $\frac{1}{2}$.  This choice is particularly appealing because it corresponds to the predictive distribution under the null model and its reference/Jeffreys prior.

To complete the specification of the LPEP we must specify the mixing distribution for the exponent $\delta$.  In this manuscript we consider three alternatives.  The first version of the LPEP we investigate is the unit information LPEP (UI-LPEP) obtained by fixing $\delta=n^*$.  We also consider 
a version of the hyper-g/n prior 
discussed in \cite{liang2008mixtures} and \cite{li2018mixtures},
\begin{align*}
f^{HGN}(\delta) &=\left(1+\frac{\delta}{n^{*}}\right)^{-2}.
\end{align*}
We call this the HGN-LPEP The median of the hyper-g/n prior is equal to $n^{*}$, and the prior places much of its mass around this value.  It can therefore be considered as a relaxation of the unit information version of the prior.  Note that, under this hyperprior, $\delta$ is independent of the model under consideration.

Finally, we consider a version of the robust prior recommended by \cite{bayarri2012criteria},
\begin{align*}
    f^{R}(\delta|\bgam)&= \frac{1}{2 (p_{\bgam}+1)^{1/2}}\frac{ (n^{*}+1)^{1/2}}{(\delta+1)^{3/2}}\mathbf{1} \left( \delta>\frac{n^{*} - p_{\bgam}}{p_{\bgam}+1} \right),
\end{align*}
which we call the R-LPEP The robust prior satisfies a number of compelling desiderata for Gaussian linear regression models.  For example, its expectation is $\mathcal{O}(n^{*})$. However, note that, unlike our previous two choices, this prior depends on the model size.  

\subsection{Properties of the LPEP prior for logistic regression}





\subsubsection{Proper prior}

Note that our choice of $m^{*}\left(\by^{*} \mid \bX \right)$ is proper, and that because of the constraint on the value of the training samples,
$$
\pi_{\bgam} \left( \bbeta_{\bgam} \mid \by^{*}, \delta\right) = 
\frac{\tilde{f}_{\bgam}\left(\by^{*} \mid \bbeta_{\bgam}, {\delta}\right) \pi_{\bgam}^N\left(\bbeta_{\bgam}\right)}{\int  \tilde{f}_{\bgam}\left(\by^{*} \mid \bbeta_{\bgam}, {\delta}\right) \pi_{\bgam}^N\left(\bbeta_{\bgam}\right) d \bbeta_{\bgam}}
$$
is also proper for every  $\bgam$.  Therefore, the LPEP prior in \eqref{eq:LPEPlogistic} is also proper for every  $\bgam$.

\subsubsection{Tail behavior}

It straightforward to see that the unit information version of the LPEP (where $\delta = n^{*}$), $\pi^{LPEP-UI}_{\bgam}(\bbeta_{\bgam})$ has Gaussian tails.  On the other hand, as the following theorem shows, the hyper-g/n and the robust versions of the LPEP have heavier (polynomial) tails in every direction.

\begin{theorem}\label{th:tails}
Let 
$$
\zeta^{HGN}(s \mid \bv, \bgam) = \left. \pi_{\bgam}^{LPEP-HGN}(\bbeta_{\bgam})\right|_{\bbeta_{\bgam} = s \bv}$$
and
$$\zeta^{R}(s \mid \bv, \bgam) = \left. \pi_{\bgam}^{LPEP-R}(\bbeta_{\bgam})\right|_{\bbeta_{\bgam} = s \bv}
$$
for any vector $\bv$ such that $\| \bv \| = 1$.  Then there exist functions $c^{HGN}_{\bgam}(\bv)$ and  $c^{R}_{\bgam}(\bv)$ such that 
$$
\lim_{s \to \infty} \frac{\zeta^{HGN}(s \mid \bv, \bgam)}{ \left( 1 + s^2/(p_{\bgam}+1) \right)^{-\frac{p_{\bgam}+2}{2}}} =  c^{HGN}_{\bgam}(\bv) < \infty
$$
and
$$
\lim_{s \to \infty} \frac{\zeta^{R}(s \mid \bv, \bgam)}{ \left( 1 + s^2/(p_{\bgam}+1) \right)^{-\frac{p_{\bgam}+2}{2}}} =  c^{R}_{\bgam}(\bv) < \infty
$$
for every direction $\bv$ and model $\bgam$.
\end{theorem}
The proof, which is presented in Appendix \ref{ap:tails}, extends results originally presented in \cite{bayarri2012criteria}.  

One important consequence of this result is that, from an estimation (rather than model selection) perspective, $\pi^{LPEP-HGN}_{\bgam}(\bbeta_{\bgam})$ and $\pi^{LPEP-R}_{\bgam}(\bbeta_{\bgam})$ are robust, in the sense of having bounded influence in the case of likelihood-prior conflict (e.g., see \citealp{andrade2006bayesian} and \citealp{andrade2011bayesian}).

\subsubsection{Intrinsic  consistency}\label{se:intrinsic}

In addition to being proper, under mild conditions the LPEP converges to a non-degenerate prior as the size of the training sample increases.  Naturally, the exact form of the intrinsic prior depends on the asymptotic regime for the covariates associated with new observations, as well as the exact prior used for $\delta$. Theorem \ref{th:intrinsicpriorconsistency} below provides a relevant example.

\begin{theorem}\label{th:intrinsicpriorconsistency}
Assume that, as $n$ (and therefore, $n^*$) grows, the covariate vectors $\bx_1, \bx_2, \ldots$ satisfy either of the following two conditions:
\begin{itemize}
    \item[(i)] If $\bx_1, \bx_2, \ldots$ forms a deterministic sequence, then $\frac{1}{n}\bX^{T}\bX \underset{n \to \infty} \longrightarrow \bfSigma$.
    
    
    \item[(ii)] If $\bx_1, \bx_2, \ldots$ are random, then they are independent and identically distributed from a distribution with mean $\mathbf{0}$ and  finite covariance matrix $\bfSigma$.
\end{itemize}

Then, the unit information ($\delta = n^{*}$), hyper-g/n and robust versions of the LPEP have proper, non-degenerate intrinsic priors of the form
\begin{align*}
\lim_{n^{*} \to \infty} \pi_{\bgam}^{LPEP-UI}(\bbeta_{\bgam}) =  \phi_{p_{\bgam}+1}
 \left( \bbeta_{\bgam} \mid  \mathbf{0} , 4  \left[ \bfSigma_{\bgam} \right]^{-1} \right) ,
\end{align*}
\begin{align*}
\lim_{n^{*} \to \infty} \pi_{\bgam}^{LPEP-HGN}(\bbeta_{\bgam}) =  \int \phi_{p_{\bgam}+1}
 \left( \bbeta_{\bgam} \mid  \mathbf{0} , 4 \delta^{*} \left[ \bfSigma_{\bgam} \right]^{-1} \right)  \left( 1 + \delta^{*}\right)^{-2} d \delta^{*} ,
\end{align*}
and
\begin{align*}
\lim_{n^{*} \to \infty} \pi_{\bgam}^{LPEP-R}(\bbeta_{\bgam}) =  \int \phi_{p_{\bgam}+1}
 \left( \bbeta_{\bgam} \mid  \mathbf{0} , 4 \delta^{*} \left[ \bfSigma_{\bgam} \right]^{-1} \right)  \frac{1}{2(p_{\bgam}+1)^{\frac{1}{2}}} \left(\frac{1}{\delta^{*}}\right)^{\frac{3}{2}} \mathbf{1}\left( \delta^{*} > \frac{1}{p_{\bgam}+1} \right) d \delta^{*} ,
\end{align*}
where $\bfSigma_{\bgam}$ denotes the square submatrix of $\bfSigma$ that includes only the rows and columns for which $\gamma_j = 1$.

\end{theorem}

A proof of this result (versions of which have been discussed in \citealp{li2018mixtures}) can be seen in Appendix \ref{ap:properties2}.  Interestingly, we note that  
these are the same intrinsic priors associated with the prior in \cite{bove2011hyper} under the same asymptotic regime for $\bX$.

\subsubsection{Model selection consistency}

Model selection consistency refers to the ability of the procedure to choose the correct model as $n\to \infty$.  Intuitively, because the amount of information in $\pi_{\bgam}^{LPEP}$ is kept approximately constant as $n^*$ increases, we would expect that the associated Bayes factors would behave asymptotically like those computed from the Bayesian Information Criteria, which have been well studied and are known to be consistent.  The following theorem, a proof of which can be seen in Appendix \ref{ap:properties1}, formalizes that intuition for the unit information prior.

\begin{theorem}\label{th:asympconsistency}
Assume that a sequence of observations $y_1, y_2, \ldots$ is generated from some model $\bgam_T \in \{ 1\}\times \{ 0,1 \}^p$ (i.e., one of the models considered by our procedure), and that $p$ is fixed.  For the unit information LPEP with $g=n$, and under appropriate regularity conditions for how the covariate vectors $\bx_{\bgam,1}, \bx_{\bgam,2}, \ldots$ are generated, we have
$$
\lim_{n \to \infty}
\Pr^{LPEP}(\bgam = \bgam_T \mid \by) = 1 .
$$
\end{theorem}

\subsubsection{Information consistency}

Information consistency refers to the behavior of the model selection criteria for fixed sample $n$, as the observed sample $\by$ becomes increasingly more ``extreme''.  In the case of logistic regression, because the sample space for $\by$ is finite for any $n$, traditional issues of information inconsistency do not arise.








\subsection{Markov chain Monte Carlo sampling}\label{se:MCMC}

The LPEP prior can be easily combined with the Polya-Gamma augmentation of \cite{polson2013bayesian} to generate an efficient Markov chain Monte Carlo algorithm for variable selection in logistic regression.  For this purpose it is convenient to re-express \eqref{eq:PEPGLM} as a hierarchical prior where 
$$
\bbeta_{\bgam} \mid \by^{*}, \delta, \bgam \sim \normal\left(  \hat{\bbeta}_{\bgam}(\by^{*}), \delta \left\{ \bH_{\bgam}(\by^{*} )\right\}^{-1} \right) ,
$$
with $y^{*} \sim m^{*}\left( y^{*} \mid \bX \right)$ and $\delta \sim f(\delta|\bgam)$.

Now, from Theorem \ref{th:existence} of \cite{polson2013bayesian}, we can write
\begin{align*}
    f_{\bgam}(\bld{y}\mid \betag) &= \prod_{i=1}^{n} \frac{\exp\left\{ y_i \bx_{\bgam,i}^T\betag \right\}}
    {1+\exp{(\bld{x}_{i,\bgam}^T\betag)}}\\
    &\propto \prod_{i=1}^n \left(\exp\left\{(y_i - 1/2) \bx_{\bgam,i}^T\betag\right\}\int_0^\infty \exp \left\{- \frac{\omega_i}{2} \left(\bx_{\bgam,i}^T\betag \right)^2 \right\}f(\omega_i \mid 1,0) d\omega_i\right) ,
\end{align*}
where $f(\omega \mid a,b)$ denotes the density of a P\`olya-Gamma random variate with parameters $a$ and $b$.  Therefore, after introducing a vector of auxiliary random variables $\bfomega = (\omega_1, \ldots,\omega_n)$, 
\begin{align}\label{eq:fullcondibetagam}
  f(\bgam, \betag,\delta \mid  \by^{*}, \bfomega)  &\propto f(\bgam) f(\delta|
  \bgam) 
  \phi_{p_{\bgam}+1} \left( \betag  \mid \hat{\bbeta}_{\bgam}(\by^{*}), \delta \bH^{-1}_{\bgam}(\by^{*}) \right)
  \phi_{n} \left( \bz  \mid \bX_{\bgam} \betag, \bfOmega^{-1} \right) ,
\end{align}
where $\bz = ((y_1 - 1/2)/\omega_1, \ldots, (y_n - 1/2)/\omega_n)^T$, $\bfOmega = \diag\left\{ \omega_1, \ldots, \omega_n \right\}$, $f(\bgam)$ is a prior on $2^p$ dimensional model space 
and $f(\delta|
  \bgam)$ is the prior on scale parameter $\delta$. 

It is straightforward to see that $\betag$ can be integrated out of \eqref{eq:fullcondibetagam}, yielding
\begin{align}\label{eq:marginallikelihood}
f(\bgam, \delta \mid \by^{*}, \bfomega) = \int f(\bgam, \betag, \delta \mid \by^{*},  \bfomega) d \betag \propto f(\bgam) f(\delta|\bgam) \phi_{n}(\bz|\bm_{\bz}^{\bgam},\bV_{\bz}^{\bgam}) ,
\end{align}
where
$\bm_{\bz}^{\bgam} = \bX_{\bgam}\hat{\bbeta}_{\bgam}(\by^{*})$, $
\bV_{\bz}^{\bgam} = \bfOmega^{-1}+ \delta\bX_{\bgam} \bH^{-1}_{\bgam}(\by^{*})\bX_{\bgam}^T$ and, as before, $\phi_{p}(\cdot \mid \bfmu, \bfSigma)$ denotes the density of the $p$-variate normal distribution with mean vector $\bfmu$ and covariance matrix $\bfSigma$.
%
%
%
Various versions of Metropolis-Hastings algorithms can be implemented to explore the space of model (e.g., see section 4.5 of \citealp{george1997approaches}).

Once the model $\bgam$ and the exponent $\delta$ have been updated, the regression coefficients can be sampled using the fact that $\betag \mid \bgam, \delta, \by^{*}, \bfomega \sim \normal \left( \bm_{\bfomega}, \bV_{\bfomega}\right)$,
where 
\begin{align}\label{eq:parambetapost}
\bm_{\bfomega} &= \bV_{\bfomega}\left(\bX_{\bgam}\bfOmega\bz+\frac{1}{\delta}\bH_{\bgam}(\by^{*})\hat{\bbeta}_{\bgam}(\by^{*})\right), &
\bV_{\bfomega} &= \left(\bX_{\bgam}^T \bfOmega\bX_{\bgam} + \frac{1}{\delta}\bH_{\bgam}(\by^{*})\right)^{-1}
\end{align}

Conditional on $\betag$, the remaining parameters $\by^{*}$ and $\bfomega$ can be easily sampled using either Gibbs sampling or random-walk Metropolis-Hastings steps.  Further details of the computational algorithm can be see in Appendix \ref{ap:MCMCMdetails}.

\section{Simulation studies}\label{se:simulation}

We conducted two simulation studies to compare the estimation and model selection performance of Laplace PEP priors with other existing model selection techniques. The setup for our simulation study is motivated by that in \cite{li2018mixtures}. This section discusses the results from the first simulation study, the results for the second one can be seen in the supplementary materials.

The simulation study described in this section uses a sample size of $n=500$ and a total number of covariates $p = p_{\bgam_{F}}=100$, with the vectors of predictors being drawn independently from a zero-mean, unit-scale multivariate normal distribution with pairwise correlations given by $cor(x_{i,j},x_{i,j'})=r^{|j-j'|}$
for $1 \leq j < j' \leq p$.  It consists of eight scenarios, which differ in terms of both the sparsity level in the vector of regression coefficients and the the correlation structure among predictors. More specifically, we consider all combinations of four different levels of sparsity ($p_{\bgam_{T}} \in \{0, 5, 10, 20 \}$, please see Table \ref{ref:coef}) and two different correlation coefficients ($r \in \{0, 0.75 \}$).
\begin{table}[!h]
    \centering
    \begin{tabular}{cccccc}
    \hline 
    $p_{\bgam_{T}}$ & $\beta_{\bgam_T,0}$ &  $\beta_{\bgam_T,1:5}$ &  $\beta_{\bgam_T,6:10}$ &  $\beta_{\bgam_T,11:15}$ &  $\beta_{\bgam_T,16:20}$\\\hline  
    0     & $-0.5$ & $\bzero$ & $\bzero$ & $\bzero$ & $\bzero$  \\
    5     &  $-0.5$ &  $\bb$ & $\bzero$ & $\bzero$ & $\bzero$\\
    10     & $-0.5$ &  $\bb$ & $\bzero$ & $\bb$ & $\bzero$\\
    20    & $-0.5$ &  $\bb$ & $0.5\bb$  & $\bb$ & $0.5\bb$ \\\hline
    \end{tabular}
    \caption{Value of intercept and coefficients in the true logistic regression model where $ \bb=(2,-1,-1,0.5,-0.5)^T$}.
    \label{ref:coef}
\end{table}

A total of 100 datasets were generated for each of our 8 scenarios.  We apply both Bayesian procedures and various penalized likelihood approaches to each dataset.  In terms of Bayesian procedures, in addition to the LPEP prior, we also consider the methodology of \citep{li2018mixtures}, which relies of a mixture of g-priors along with a Laplace approximation to compute the associated marginal likelihood (denoted LCL in the sequel), as well as an ``exact'' version of their procedure that relies on the same mixture of g-priors but avoids the Laplace approximation by implementing a latent-variable augmentation similar to the one described in Section \ref{se:MCMC} (denoted LCE in the sequel). Comparing LCL and LCE allows us to disentangle the effect of the Laplace approximation from that of the prior choice on the performance of these techniques.  For each of these three approaches, we consider three different settings for the hyperparameter $\delta$:  the unit information prior with $\delta = n$, the hyper-g/n and robust priors (recall Section \ref{se:LPEPlogistic}). We use the \texttt{R} package \texttt{BAS} \citep{BAS} to implement LCL, and a slight modification of our own code to implement LCE.  In all cases we assume a Beta-Binomial(1,1) prior over the model space, and run the MCMC chain for $2^{17}\approx 131,000$ iterations after a burn-in of $10,000$ iterations.  Note that we do not include the CRPEP and DRPEP priors from \cite{fouskakis2018power} in this simulation study.  We do this for two main reasons.  First, the computational complexity of the code provided by the authors makes a simulation study like this prohibitive.  Not only is each iteration of the algorithm much more expensive than those of the other approaches, but the algorithm mixes much more slowly, which means that a much larger number of iterations are required to get accurate results.  Secondly, and more important, the algorithm we had access to broke down for a number of our simulated datasets.  We incorporate comparisons with CRPEP and DRPEP in our real data examples in Section \ref{se:realdata}.  In terms of penalized likelihood methods, we compare against LASSO \citep{tibshirani1996regression}, smoothly clipped absolute deviation (SCAD) \citep{fan2001variable} and minimax concave penalty (MCP) \citep{zhang2010nearly}.  We use the \texttt{R} package \texttt{glmnet} \citep{friedman2010regularization} to implement LASSO, and the package \texttt{ncvreg} \citep{ncvreg} for SCAD and MCP.

We first evaluate the performance of these various methods in terms of model selection performance using three metrics.  First, we report the frequency (over the 100 datasets) with which the MAP model matches the true model $\bgam_T$ (see Table \ref{ref: MAP}).  For the penalized likelihood approaches (for which a single model is reported for each dataset) the equivalent metric is simply the number of datasets for which the technique reported the correct model. This metric was used in both \cite{li2018mixtures} and \cite{fouskakis2018power}.  

\begin{table}[!h]
    \centering
\scalebox{1}{
    \begin{tabular}{c|c|cccccccc}
    \toprule
    \multicolumn{2}{c}{$p$} & \multicolumn{8}{c}{100}  \\
    \hline
    \multicolumn{2}{c}{$p(\bgam)$} &
     \multicolumn{8}{c}{Beta-Binomial(1,1)}  \\
  \hline
    \multicolumn{2}{c}{$p_{\bgam_{T}}$} &
     \multicolumn{2}{c}{0} & \multicolumn{2}{c}{5} & \multicolumn{2}{c}{10} & \multicolumn{2}{c}{20}   \\
  \cmidrule(lr){1-2}\cmidrule(lr){3-4}\cmidrule(lr){5-6} \cmidrule(lr){7-8}\cmidrule(lr){9-10}
     \multicolumn{2}{c}{$r$} & 0 & 0.75 & 0 & 0.75& 0 & 0.75& 0 & 0.75  \\ \midrule
\multirow{ 3}{*}{$\delta=n$} & LPEP & 99 & \textbf{100}* & 45 & 4 & \textbf{18}* & 0 & 0 & 0 \\ 
  & LCE & \textbf{100}* & \textbf{100}* & 45 & \textbf{5} & 8 & 0 & 0 & 0 \\
    & LCL & \textbf{100}* & \textbf{100}* & \textbf{46} & 4 & 11 & 0 & 0 & 0 \\ 
\midrule
 \multirow{ 3}{*}{$\delta\sim \text{robust}$} & LPEP & 99 & \textbf{100}* & \textbf{53}* & \textbf{6}* & \textbf{15} & 0 & 0 & 0 \\ 
  & LCE & 99 & \textbf{100}* & 45 & \textbf{6}* & 0 & 0 & 0 & 0 \\ 
    & LCL & \textbf{100}* & \textbf{100}* & 46 & \textbf{6}* & 2 & 0 & 0 & 0 \\ 

  \midrule
  \multirow{ 3}{*}{$\delta\sim \text{hyper g/n}$}& LPEP & \textbf{98} & \textbf{100}* & \textbf{50} & \textbf{5} & \textbf{17} & 0 & 0 & 0 \\ 
  & LCE & 97 & 99 & 25 & 4 & 0 & 0 & 0 & 0 \\ 
  & LCL & 65 & 78 & 3 & 0 & 0 & 0 & 0 & 0 \\ 
  \midrule
  & LASSO & 59 & 65 & 0 & 0 & 0 & 0 & 0 & 0 \\ 
  & SCAD & 57 & 59 & 0 & 0 & 0 & 0 & 0 & 0 \\ 
  & MCP & \textbf{73} & \textbf{66} & \textbf{8} & 0 & \textbf{3} & 0 & 0 & 0 \\ \bottomrule
    \end{tabular}}
    \caption{Number of times (\textbf{over 100 replications}) that the \textbf{MAP} model coincides with the true model in the logistic regression ; \textbf{BOLD} represent group maximum; * represent overall maximum. }
    \label{ref: MAP}
\end{table}

First note that Bayesian methods tend to clearly outperform penalized likelihood approaches, in some cases quite dramatically.  Focusing now on the Bayesian approaches, we observe that most of them perform very well when the data is generated from the null model.  This is true both for uncorrelated and highly correlated covariates.  The main exception is LCL under the hyper-g/n hyperprior, where the MAP algorithm matches the true model in only 65 ($r=0$) and 78 ($r=0.75$) of the datasets.  On the other hand, as the number of non-zero coefficients in the true model increases, we observe that all approaches struggle to identify the true model, particularly when the covariates are highly correlated.  In particular, when $p_{\bgam_{T}} = 20$, none of the procedures is able to identify the true model.  Nonetheless, it appears that, overall, LPEP (and, specially, the robust and the hyper-g/n versions of LPEP) perform the best.

\begin{figure}[!ht]
    \centering
    \includegraphics[width=0.9\linewidth,height=0.9\textheight]{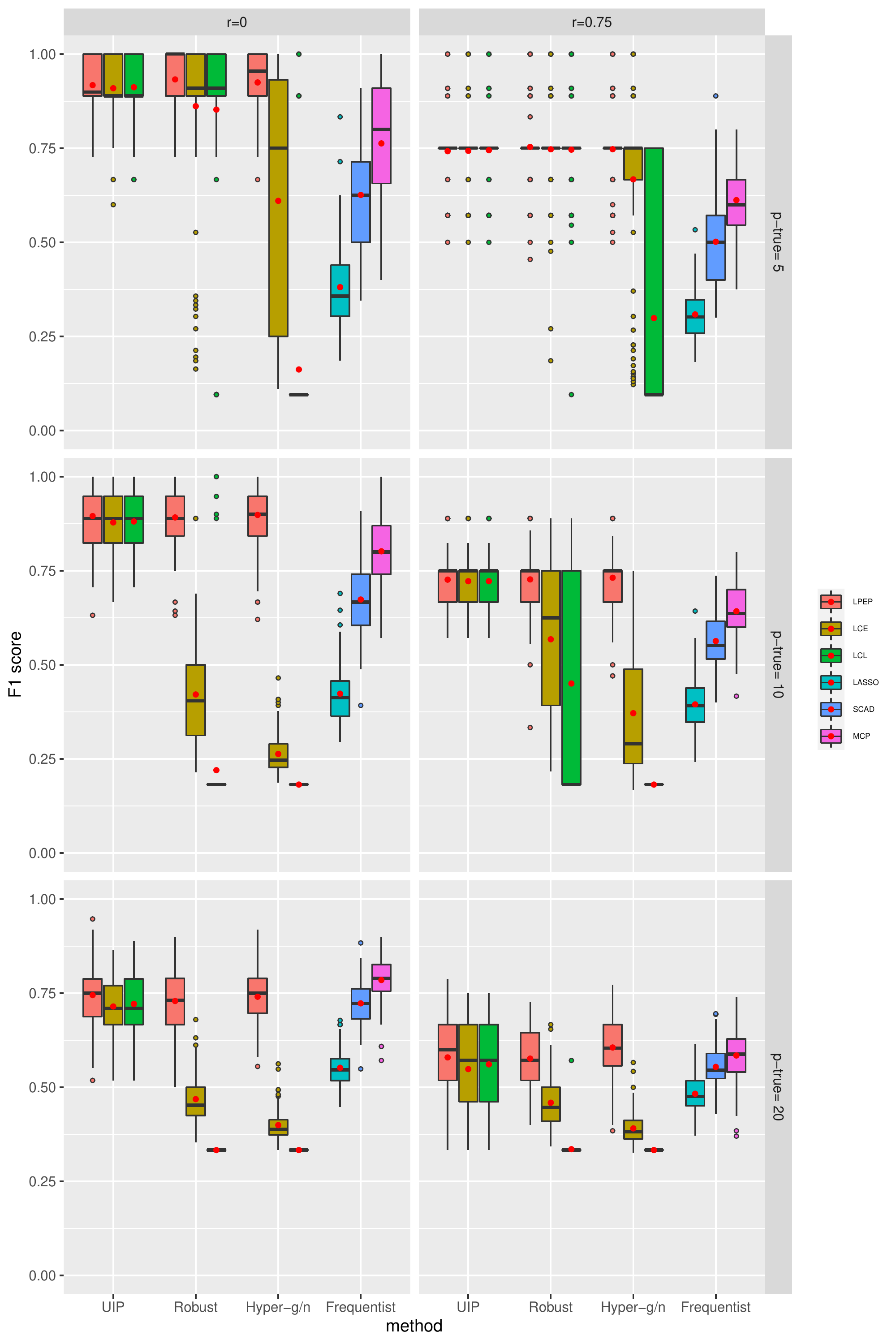}
    \caption{F1 score for the MAP model estimated by various methods and prior combinations for 100 simulated datasets ($n=500,p=100$) under different scenarios of correlation ($r=0$: left; $r=0.75$: right) and true number of non-zero coefficients specified in rows ($p-true=p_{\bgam_{T}}$); Red dots represent the average F1 score across 100 simulated datasets.}
    \label{fig:F1score}
\end{figure}

While the MAP metric we discussed above provides some insights into model performance, it tends to be less informative when there is substantial uncertainty on the posterior distribution over the model space.  Therefore, we also compute for each dataset the  $F_1$ score for the MAP (Bayesian procedures) or selected (penalized likelihood procedures) model, see Figure \ref{fig:F1score}.  In this setting, the $F_1$ score is defined as the harmonic mean of proportion of true positives among ``selected” covariates (the precision) and the proportion of ``selected” covariates among true positive covariates (the recall). The $F_1$ score ranges between 0 and 1, with a higher value indicating better model selection performance. Note that results are not presented for the null model since the $F_1$ score is not well defined in that case.

The $F_1$ score provides a much more informative picture of the performance of these models.  In all cases, the methods based on LPEP priors tend to perform the best, with the robust and hyper-g/n versions being slightly better than that of the unit information prior.  We also see that, while all Bayesian procedures have very similar performance under the unit information prior, LCE tends to outperform LCL under the robust and hyper-g/n priors (in some cases, quite dramatically).

\begin{figure}[ht]
    \centering
    \includegraphics[width=0.9\linewidth,height=0.9\textheight]{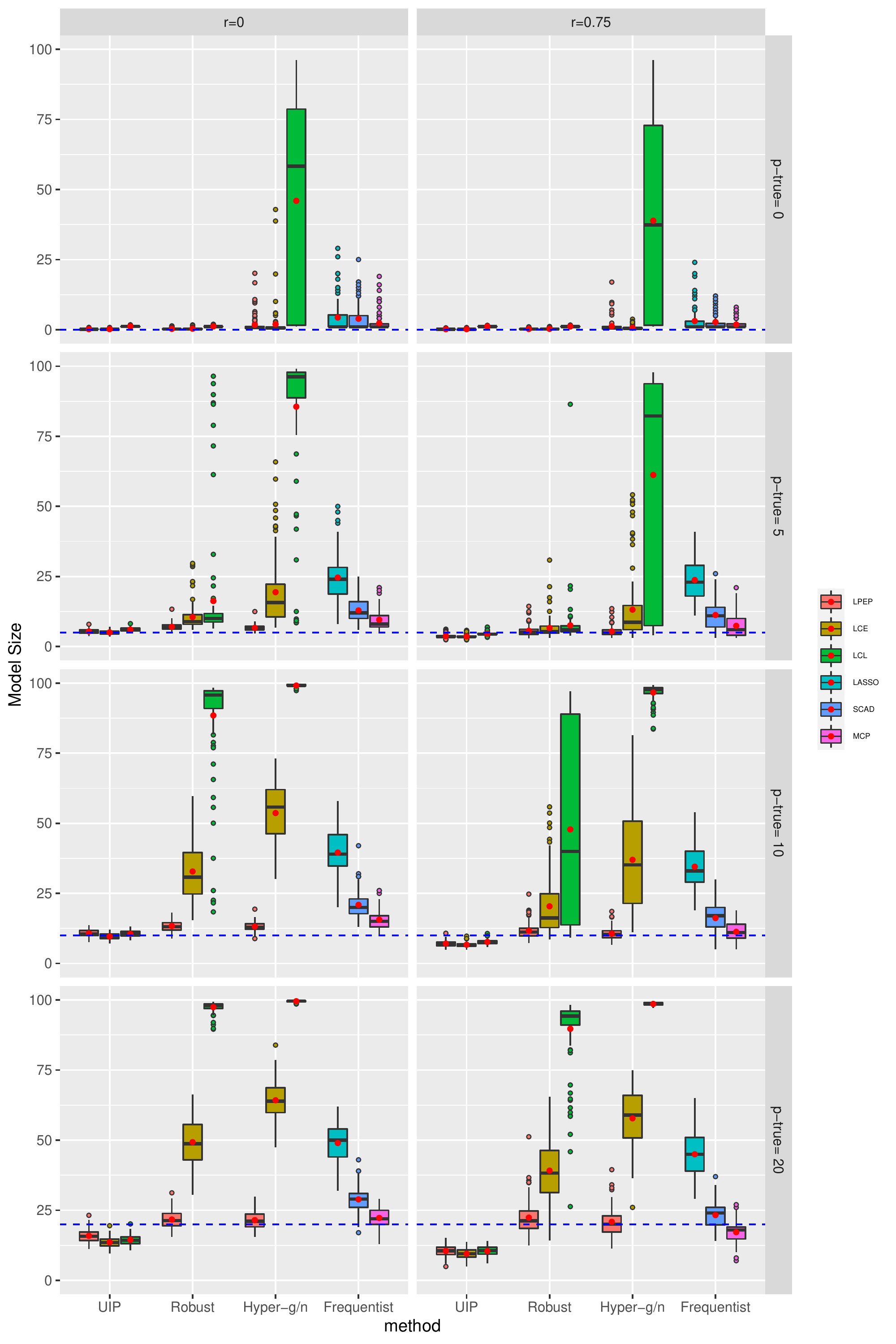}
    \caption{Average size of models selected by various methods and prior combinations for 100 simulated datasets ($n=500,p=100$) under different scenarios of correlation ($r=0$: left; $r=0.75$: right) and true number of non-zero coefficients specified in rows ($p-true=p_{\bgam_{T}}$); Dotted blue line indicates the true model size $p-true=p_{\bgam_{T}}$ and red dots represent the average model size over 100 simulated datasets.}
    \label{fig:modsize}
\end{figure}

Next, we also report the average size of the sampled/selected models for each data set (see Figure \ref{fig:modsize}).  Under the robust and hyper-g/n priors, LCE and, especially, LCL tend to favor very large models.  Interestingly, all Bayesian procedures under the unit information prior seem to underestimate the model size when $p_{\bgam_T} = 20$.  The best performing approaches with average model size close to true model size are again the robust and hyper-g/n versions of LPEP.

Finally, we compare the procedures in terms of parameter estimation performance using the average mean squared error (AMSE) of the estimated coefficients 
$$
AMSE(\bbeta)=\frac{1}{p}\sum_{j=1}^p(\hat{\beta}_j-\beta_{j,\bgam_T})^2
$$
where $\hat{\beta}_j$ and $\beta_{j,\bgam_T}$ are the estimated and true values of $j^{th}$ covariate, respectively. For the Bayesian procedures, model-averaged posterior mean estimates are used for this calculation. For penalized likelihood methods, the sparse point estimates of the coefficients are used. The results can be seen in Table \ref{ref:MSE}.

\begin{table}[h]
    \centering
\scalebox{1}{
    \begin{tabular}{c|c|cccccccc}
    \toprule
    \multicolumn{2}{c}{$p$} & \multicolumn{8}{c}{100}  \\
    \hline
    \multicolumn{2}{c}{$p(\bgam)$} &
     \multicolumn{8}{c}{Beta-Binomial(1,1)}  \\
  \hline
    \multicolumn{2}{c}{$p_{\bgam_{T}}$} &
     \multicolumn{2}{c}{0} & \multicolumn{2}{c}{5} & \multicolumn{2}{c}{10} & \multicolumn{2}{c}{20}   \\
  \cmidrule(lr){1-2}\cmidrule(lr){3-4}\cmidrule(lr){5-6} \cmidrule(lr){7-8}\cmidrule(lr){9-10}
     \multicolumn{2}{c}{$r$} & 0 & 0.75 & 0 & 0.75& 0 & 0.75& 0 & 0.75  \\ \midrule
\multirow{ 3}{*}{$\delta=n$} & LPEP & 0.11 & \textbf{0.10}* & 2.91 & \textbf{7.67} & 7.09 & \textbf{17.67} & \textbf{14.70} & \textbf{33.90} \\ 
& LCE & 0.11 & \textbf{0.10}* & 3.06 & 7.78 & 7.64 & 18.44 & 16.11 & 36.47 \\  &  LCL & \textbf{0.10}* & \textbf{0.10}* & \textbf{2.87} & 7.68 & \textbf{6.78} & 18.17 & 16.22 & 36.43 \\ 
 \midrule
 \multirow{ 3}{*}{$\delta\sim \text{robust}$} & LPEP & 0.12 & \textbf{0.10}* & \textbf{2.62}* & \textbf{6.87}* & \textbf{6.04}* & \textbf{14.07}* & \textbf{13.38} & \textbf{24.03}* \\ 
 & LCE & 0.12 & 0.11 & 4.83 & 7.80 & 47.30 & 23.30 & 96.14 & 52.80 \\ 
 & LCL & \textbf{0.10}* & \textbf{0.10}* & 8.86 & 8.44 & 214.63 & 60.56 & 275.93 & 115.58 \\ 
 \midrule
  \multirow{ 3}{*}{$\delta\sim \text{hyper g/n}$} & LPEP & \textbf{0.16} & 0.14 & \textbf{2.70} & \textbf{6.89} & \textbf{6.12} & \textbf{14.76} & \textbf{13.03}* & \textbf{24.86} \\ 
 & LCE & 0.23 & \textbf{0.13} & 6.71 & 8.90 & 38.54 & 26.25 & 51.48 & 44.29 \\  & LCL & 0.29 & 0.31 & 34.28 & 22.93 & 104.10 & 72.95 & 130.80 & 94.98 \\ 
\midrule
 & LASSO & 0.25 & 0.20 & 7.08 & 11.91 & 17.15 & 25.04 & 29.44 & 36.69 \\ 
 & SCAD & \textbf{0.21} & \textbf{0.16} & 3.07 & 9.02 & 6.62 & \textbf{18.80} & 14.88 & \textbf{33.00} \\ 
 & MCP & 0.22 & 0.18 & \textbf{2.82} & \textbf{8.92} & \textbf{6.35} & 19.38 & \textbf{15.13} & 33.52 \\  \bottomrule
    \end{tabular}}
    \caption{1000 times the AMSE for estimated coefficients over 100 replications; \textbf{BOLD} represent group minimum; * represent overall minimum.}
    \label{ref:MSE}
\end{table}

We observe that as the true model size $p_{\bgam_T}$ and the true correlation between covariates increases, the AMSE increases for all techniques. However, similar to model selection performance, LPEP versions significantly outperforms all other techniques in terms of estimation performance under non-null true model scenarios and is comparable to other techniques when the true model is the null model.

\section{Real data applications}\label{se:realdata}

\subsection{\texttt{URINARY}: Determinants of urinary incontinence}\label{se:urinary}

The \texttt{URINARY} data set describes the results from a small drug study with 21 subjects.  The response corresponds to whether the subject developed urinary incontinence after receiving the drug.  The explanatory variables capture drug-induced physiological changes, which were in the same direction for most subjects.  This data set was first presented in  \cite{potter2005permutation}, and is further discussed in \cite{mansournia2018separation}.  While very small, the data set is challenging to analyze because it exhibits full separation.  In particular, the maximum likelihood estimates of the regression coefficients are all infinite (please see the top row of Table \ref{tab:urinarycoef}), indicating that separation is not induced by any of the variables on its own, but by a non-trivial linear combination of them.

In addition to the maximum likelihood estimates reported by the \texttt{R} function \texttt{glm}, Table \ref{tab:urinarycoef} presents estimates for the regression coefficients for various Bayesian and penalized likelihood methods. The results for LPEP, CRPEP and DRPEP are based on 10,000 iterations of the MCMC algorithm obtained after a burn-in period of 10,000 iterations. On the other hand, for LCL we use full model enumeration procedure in the \texttt{R} package \texttt{BAS}. In the case of Bayesian procedures, Table \ref{tab:urinarycoef} presents model-averaged posterior means, as well as 95\% model-averaged credible intervals for the coefficients.  Note that confidence intervals for the penalized likelihood procedures are not presented since they are not straightforward to obtain and the \texttt{R} packages we used to fit these models do not readily provide them.  Furthermore, results for CRPEP and DRPEP are not included under the robust hyper-prior because such a procedure is not implemented in \cite{fouskakis2018power}.

Note that LCL produces large point estimates and very wide credible for the model coefficients under all hyperpriors.  This is no surprise; the prior proposed by \cite{li2018mixtures} is proper only for models for which the maximum likelihood estimates are finite.  This means that, for a data set like \texttt{URINARY}, some of the Bayes factors associated with LCL are ill-defined.  This is also the reason why we do not show results for LCE; the target posterior distribution for the associated Markov chain Monte Carlo algorithm is improper if the full model is included in the analysis.  Furthermore, note that the point estimates associated with CRPEP appear to be different from those generated by LPEP, DRPEP, and the penalized likelihood methods.  This is clearer when looking at the intercept of the model, which is negative with high probability under CRPEP but positive with high probability under LPEP and DRPEP under all hyperpriors.

\begin{table}[!h]
    \centering\scalebox{0.8}{
    \begin{tabular}{c|c|cccc}
    \toprule
 \multicolumn{2}{c}{} & $\beta_0$ & $\beta_1$& $\beta_2$ & $\beta_3$\\\midrule
 & \multirow{2}{*}{MLE} & -83.84 & -2445.04 & -1653.76 & 310.27  \\
 & & {\footnotesize (-1969.33 , 484.67)} & 
 {\footnotesize (-53259.42 , 10866.11)} & 
 {\footnotesize (-34653.95 , 5062.34)}& 
 {\footnotesize (-1166.03 , 6803.89)}\\\midrule
 \multirow{ 8}{*}{$\delta=n$} & \multirow{2}{*}{LPEP} & 0.56  & -0.70   & -0.39   & 0.15  \\
 & & {\footnotesize (-1.66 , 2.85)} & {\footnotesize(-2.32 , 0.10)} & {\footnotesize(-0.81 , -0.13)} &  {\footnotesize(0.00 , 0.37)}\\\cmidrule{3-6}
 & \multirow{2}{*}{LCL} & -83.84  & -2333.88  & -1578.58   & 296.17 \\
 & & {\footnotesize(-6009.26 , 5897.13)} & {\footnotesize(-161488.87 , 158312.76)} & {\footnotesize(-109266.01 , 107118.14)} &  {\footnotesize(-19896.81 , 20678.41)}\\\cmidrule{3-6}
 & \multirow{2}{*}{CRPEP} & -1.15 & -0.70 & -0.34 & 0.00  \\
 & & {\footnotesize (-3.21 , 0.52)} & 
 {\footnotesize(-1.88 , 0.30)} & 
 {\footnotesize(-0.63 , -0.10)} &  {\footnotesize(-0.00 , 0.00)}\\\cmidrule{3-6}
 & \multirow{2}{*}{DRPEP} & 0.69 & -1.00 & 0.00 & 0.06 \\
 & & {\footnotesize (-0.55 , 2.13)} & 
 {\footnotesize(-2.09 , -0.19)} & 
 {\footnotesize(0.00 , 0.00)} &  
 {\footnotesize(-0.03 , 0.16)}\\
\midrule
\multirow{ 4}{*}{$\delta\sim \text{robust}$} & \multirow{2}{*}{LPEP}  & 0.71  & -0.98  & -0.52  & 0.19 \\ 
& & {\footnotesize(-1.74 , 3.55)} & {\footnotesize(-3.82 , 0.05)} & {\footnotesize(-1.89 , -0.12)} & {\footnotesize(0.00 , 0.54)} \\
\cmidrule{3-6}
 & \multirow{2}{*}{LCL}  & -83.84  & -2148.67  & -1453.30  & 272.66 \\
 & & {\footnotesize(-6250.14 , 5980.74)} & {\footnotesize(-161065.38 , 154146.29)} & {\footnotesize(-108979.51 , 104298.98)} &   {\footnotesize(-19890.08 , 20102.78)}\\
  \midrule
\multirow{8}{*}{$\delta\sim \text{hyper g/n}$} & \multirow{2}{*}{LPEP}  &0.61  & -0.75  & -0.41   & 0.15    \\ 
& & {\footnotesize(-1.65 , 3.07)} & {\footnotesize(-2.74 , 0.10)} & {\footnotesize(-1.02 , -0.09)} & {\footnotesize(0.00 , 0.39)}\\
\cmidrule{3-6}
 & \multirow{2}{*}{LCL}  & -83.84  & -1288.82  & -871.72  & 163.55 \\ 
 & & {\footnotesize(-6252.44 , 5650.45)} & {\footnotesize(-124412.67 , 113166.12)} & {\footnotesize(-84179.77 , 76570.77)} &  {\footnotesize(-15457.94 , 14685.16)}\\ \cmidrule{3-6}
  & \multirow{2}{*}{CRPEP} &  -1.04 & -0.66 & -0.33 & 0.00\\
 & & {\footnotesize (-3.04 , 0.62)} & 
 {\footnotesize(-1.78 , 0.30)} & 
 {\footnotesize(-0.65 , -0.08)} &  {\footnotesize(-0.00 , 0.00)}\\\cmidrule{3-6}
 & \multirow{2}{*}{DRPEP} & -0.89 & 0.00 & -0.36 & 0.00 \\
 & & {\footnotesize (-3.11 , 0.69)} & 
 {\footnotesize(0.00 , -0.00)} & 
 {\footnotesize(-0.76 , -0.11)} &  
 {\footnotesize(-0.00 , 0.00)}\\
  \midrule
 & LASSO & 0.36 & -0.70 & -0.31 & 0.11\\ 
 & SCAD &   0.41 & -0.23 & -0.20 & 0.07\\ 
 & MCP & 0.40 & -0.17 & -0.20 & 0.07 \\ \bottomrule
    \end{tabular}}
    \caption{Estimated BMA coefficients and 95\% credible intervals for  different Bayesian techniques for \texttt{urinary} dataset; For frequentist techniques, estimated coefficient is displayed.}
    \label{tab:urinarycoef}
\end{table}

Next, we present in Table \ref{tab:urinarymodel} the posterior probabilities associated with each of the eight models under consideration under each one of the Bayesian approaches, along with the model selected by each of the penalized likelihood methods.  In all cases, LCL consistently places probability one on the full model, which is also the model selected by all the penalized likelihood methods.  The full model is also consistently preferred by LPEP, but there is more uncertainty. Indeed, under LPEP the model that excludes the first covariate receives between 0.19 and 0.22 probability, and the model that excludes the third covariate is assigned between 0.03 and 0.04 posterior probability.  In contrast, CRPEP and DRPEP place zero probability on the full model. Instead, CRPEP consistently favors the model that excludes the third covariate, while DRPEP yields contradictory results depending on the hyperprior: it favors the model that excludes the second variable under the $g=n$ hyperprior, but the model that only includes the second variable under the hyper-g/n prior.

\begin{table}[!h]
    \centering\scalebox{0.9}{
    \begin{tabular}{c|c|cccccccc}
    \toprule
\multicolumn{2}{c}{} & \multicolumn{8}{c}{$(\gamma_1,\gamma_2,\gamma_3)$}\\\cmidrule{3-10}
 \multicolumn{2}{c}{} & $(0,0,0)$ & $(1,0,0)$ & $(0,1,0)$ &$(1,1,0)$ &$(0,0,1)$ &$(1,0,1)$ &$(0,1,1)$ &$(1,1,1)$  \\\midrule
 \multirow{ 4}{*}{$\delta=n$} & LPEP & 0.00 & 0.00 & 0.06 & 0.03 & 0.00 & 0.00 & 0.22 & \textbf{0.69} \\ 
 & LCL & 0.00 & 0.00 & 0.00 & 0.00 & 0.00 & 0.00 & 0.00 &\textbf{1.00}\\ 
 & CRPEP & 0.00 & 0.00 & 0.00 & \textbf{1.00} & 0.00 & 0.00 & 0.00 & 0.00 \\
 & DRPEP & 0.00 & 0.00 & 0.00 & 0.00 & 0.00 & \textbf{1.00} & 0.00 & 0.00\\
\midrule
\multirow{ 2}{*}{$\delta\sim \text{robust}$} & LPEP & 0.00 & 0.00 & 0.06 & 0.03 & 0.00 & 0.00 & 0.19 & \textbf{0.71} \\ 
 & LCL & 0.00 & 0.00 & 0.00 & 0.00 & 0.00 & 0.00 & 0.00 & \textbf{1.00} \\ 
  \midrule
\multirow{4}{*}{$\delta\sim \text{hyper g/n}$} & LPEP & 0.00 & 0.00 & 0.06 & 0.04 & 0.00 & 0.00 & 0.20 & \textbf{0.69} \\ 
 & LCL & 0.00 & 0.00 & 0.00 & 0.00 & 0.00 & 0.00 & 0.00 & \textbf{1.00} \\ 
  & CRPEP & 0.00 & 0.00 & 0.00 & \textbf{1.00} & 0.00 & 0.00 & 0.00 & 0.00\\
 & DRPEP & 0.00 & 0.00 & \textbf{1.00} & 0.00 & 0.00 & 0.00 & 0.00 & 0.00 \\
  \midrule
  & LASSO & 0.00 & 0.00 & 0.00 & 0.00 & 0.00 & 0.00 & 0.00 &\textbf{1.00} \\ 
   & SCAD & 0.00 & 0.00 & 0.00 & 0.00 & 0.00 & 0.00 & 0.00 & \textbf{1.00} \\ 
   & MCP & 0.00 & 0.00 & 0.00 & 0.00 & 0.00 & 0.00 & 0.00 & \textbf{1.00} \\  \bottomrule
    \end{tabular}}
    \caption{Posterior model probabilities for all the models in the model space for \texttt{urinary} dataset.}
    \label{tab:urinarymodel}
\end{table}

\subsection{\texttt{GUSTO-I}: Survival to treatments for occluded coronary arteries}

Next, we consider data from the  Global Utilization of Streptokinase and TPA for Occluded Coronary Arteries (GUSTO-I) trial \citep{califf1996one}, which has been previously analyzed in \cite{held2015approximate} and \cite{li2018mixtures}, and is publicly available at \url{http://www.clinicalpredictionmodels.org/} \citep{steyerberg2019clinical}. Similar to previous analyses, we model the binary endpoint of 30-day survival for a subgroup of $n=2188$ patients using 17 clinical covariates described in the supplementary materials.  

Figure \ref{fig:gustopip} displays the  marginal posterior inclusion probabilities (PIPs) for Bayesian methods and the inferred model under the penalized likelihood techniques. For techniques related to LCL, we again rely on full model enumeration.  On the other hand, for all other Bayesian techniques, we use 131,000 iterations with a burn-in of 10,000 iterations. 

Similar to simulation studies, all penalized likelihood techniques select denser models than the Bayesian procedures. In line with \cite{li2018mixtures} and \cite{held2015approximate}, we observe that \texttt{AGE}, \texttt{KILLIP}, \texttt{HYP}, \texttt{HRT} and \texttt{STE} have high PIPs under all methods. However, it is worthwhile noting that the different versions of LCL perform quite differently. In particular, the version of LCL that relies on a hyper-g/n hyperprior tends to explore very dense models leading, to PIPs close to $0.5$ for all variables. 
Similarly, the hyper-g/n versions of CRPEP and DRPEP seem to differ from their $g=n$ versions with respect to \texttt{PMI} and \texttt{SEX} variables.
On the other hand, the different versions of the LPEP prior are roughly in agreement for all variables

\begin{figure}[!h]
    \centering
    \includegraphics[width=\textwidth]{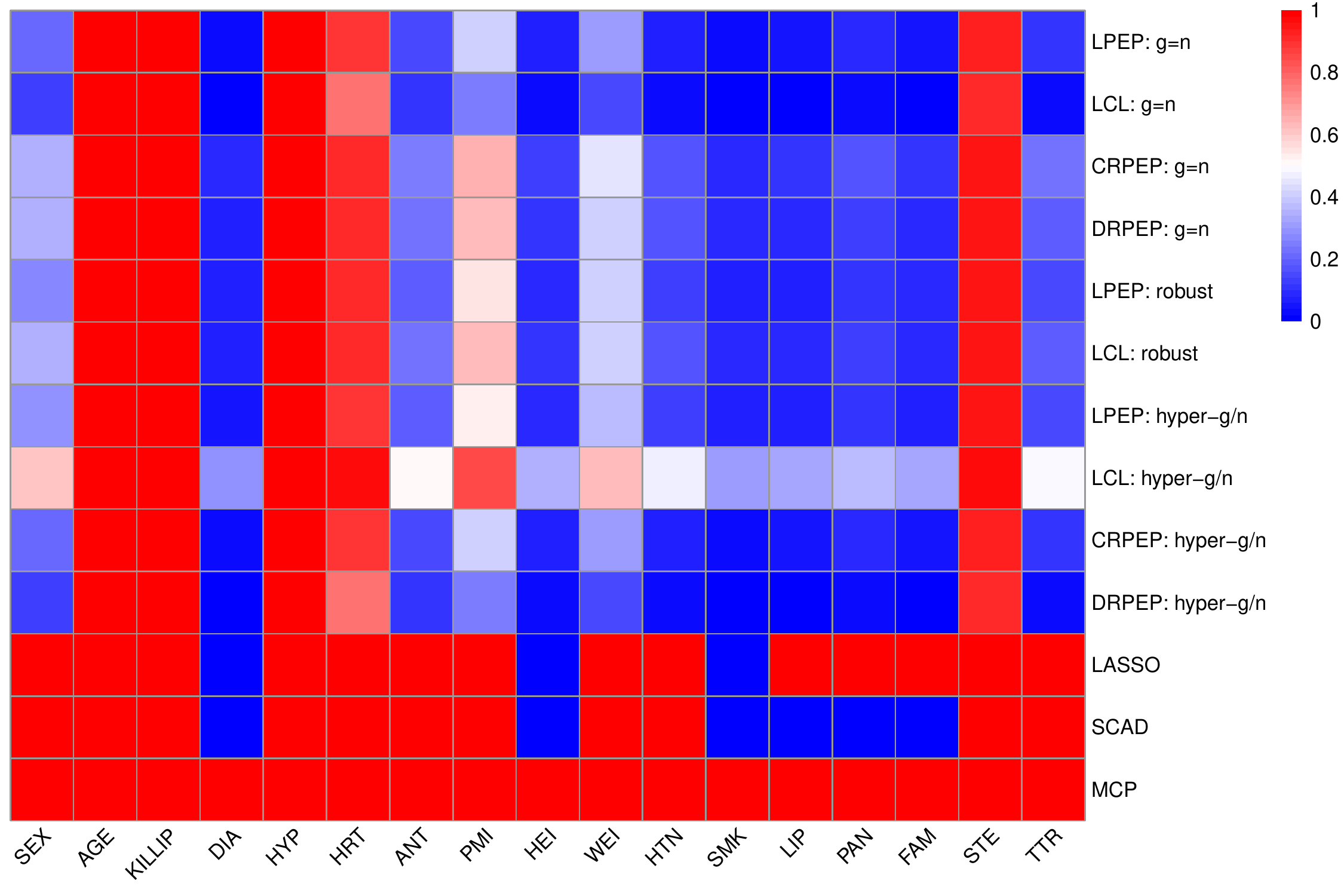}
    \caption{Marginal posterior inclusion probabilities (PIPs) for \texttt{GUSTO-I} dataset (Bayesian procedures) and variables included in the model (penalized likelihood methods).}
    \label{fig:gustopip}
\end{figure}

As in \cite{li2018mixtures}, we also compare the different procedures in terms of their out-of-sample predictive performance.  For this purpose, we performed a 10-fold cross-validation study. More specifically, we divided the data into 10 folds, training our model on 9 them and evaluating the predictive accuracy on the remaining one.  Table \ref{tab:gustopred} presents the average value of four different metrics across all 10 folds.  The metrics we rely on are the same ones employed in \cite{li2018mixtures}: the area under the ROC curve (AUC), the Calibration Slope (CS), the Logarithmic Score (LS) and the Brier score (BRIER).  AUC and CS allow us to  evaluate the methods in terms of discrimination and calibration.  In both cases, scores closer to 1 indicate better performance. On the other hand, LS and BRIER measure the predictive accuracy of methods; in both cases lower scores indicate better performance. 

Table \ref{tab:gustopred} suggests that most methods perform similarly.  The main exceptions are both versions of CRPEP and DRPEP, which seem to substantially underperform across all metrics. LPEP procedures slightly outperforming other methods in terms of AUC and CS. On the other hand, LASSO seems to slightly outperform the LPEP procedures in terms of LS and Brier score, but at the cost of selecting much denser models.   

\begin{table}[!h]
    \centering\scalebox{0.9}{
    \begin{tabular}{c|c|cccc}
    \toprule
\multicolumn{2}{c}{} & AUC & CS & LS & BRIER \\\midrule

 \multirow{ 4}{*}{$\delta=n$} & LPEP & \textbf{0.8324}* & 0.9971 & \textbf{0.1824} & \textbf{0.0496} \\ 
 & LCL & 0.8300 & 0.9931 & 0.1831 & 0.0497 \\ 
 & CRPEP & 0.7789 & 1.0578 & 0.1965 & 0.0521 \\
 & DRPEP & 0.7790 & 1.0569 & 0.1963 & 0.0521\\
\midrule
\multirow{ 2}{*}{$\delta\sim \text{robust}$} & LPEP & \textbf{0.8322} & \textbf{1.0129} & \textbf{0.1822} & \textbf{0.0495} \\ 
 & LCL &  0.8316 & 0.9804 & \textbf{0.1822} & \textbf{0.0495}\\ 
  \midrule
\multirow{4}{*}{$\delta\sim \text{hyper g/n}$} & LPEP & \textbf{0.8319} & \textbf{1.0074}* & 0.1823 & 0.0495 \\ 
 & LCL & 0.8311 & 1.0109 & \textbf{0.1818} & \textbf{0.0493}\\ 
  & CRPEP & 0.7956 & 1.1677 & 0.1951 & 0.0522 \\
 & DRPEP & 0.7800 & 1.0571 & 0.1961 & 0.0520 \\
  \midrule
  & LASSO & \textbf{0.8305} & \textbf{1.0369} & \textbf{0.1816}* & \textbf{0.0492}* \\ 
   & SCAD & 0.8243 & 0.9135 & 0.1838 & 0.0496   \\ 
   & MCP & 0.8250 & 0.9196 & 0.1838 & 0.0496\\  \bottomrule
    \end{tabular}}
    \caption{Average prediction accuracy measures in a 10-fold cross validation study for \texttt{GUSTO-I} dataset; Bold represents group maximum for AUC, for CS closest to one, and group minimum for LS and Brier score; * represents the corresponding best score among all methods.}
    \label{tab:gustopred}
\end{table}

\subsection{\texttt{HOUSE107}:  Determinants of legislator behavior in 107\textsuperscript{th} U.S. House of Representatives}

Most analyses of congressional voting treat all roll-call votes in the same way, independently of the type of vote. This might mask considerable variation in voting behavior across different types of votes.  For example, \cite{jessee2012two} provide empirical evidence that the forces affecting legislators’ voting on procedural and final passage matters have exhibited important changes over time, with differences between these two vote types becoming larger, particularly in recent congresses.  

One shortcoming of the methodology presented in \cite{jessee2012two} is that it provides legislature-wide measures of agreement among vote types, but cannot ascribe observed differences to individual legislators. Recently, \cite{lofland2017assessing} and \cite{moser2021multiple} developed methodology that enables the identification of differences in voting behavior across votes types for individual legislators.  In this section we analyze a dataset where the response variable corresponds to estimates of whether each legislator in the 107\textsuperscript{th} U.S.\ House of Representatives share the same voting behavior across final passage, amendment and procedural votes.  These estimates are obtained using the model introduced in \cite{moser2021multiple}.  Hence, in this case, $y_i = 1$ if the $i$-th legislator voting preferences remain unchanged across all three vote types, and $y_i = 0$ otherwise. The goal of this analysis is to understand whether a group of 26 characteristics of the legislator or its constituency affect the likelihood of such changes. Linking voting behavior with these underlying characteristics can provide important insights into the workings of a political system (e.g., see \citealp{facchini2011drives},  \citealp{o2011determinants} and \citealp{cragg2013carbon}).

\begin{figure}[!h]
    \centering
    \includegraphics[width=\textwidth]{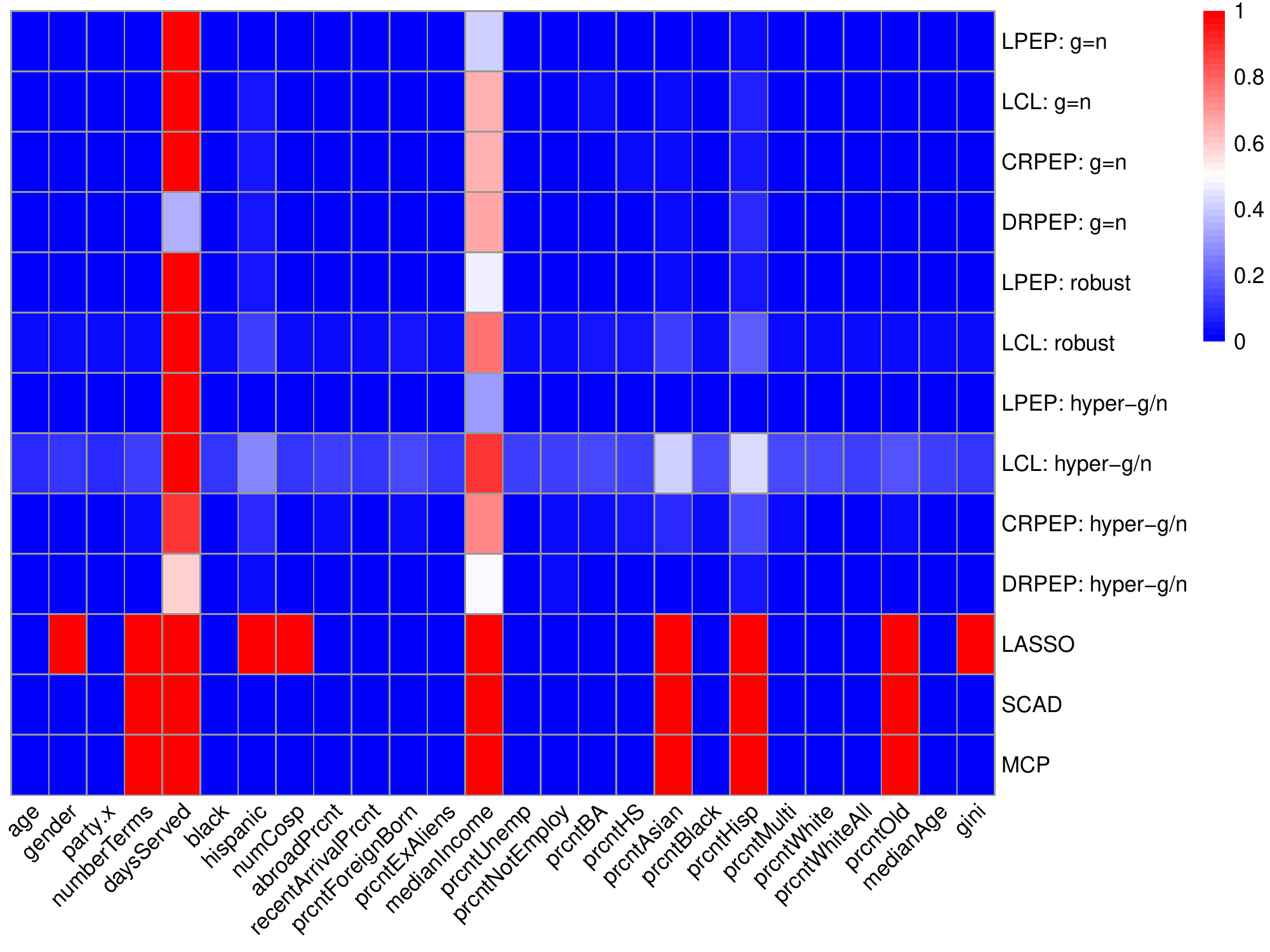}
    \caption{Marginal posterior inclusion probabilities (PIPs) for \texttt{HOUSE107} dataset (Bayesian procedures) and variables included in the model (penalized likelihood methods).}
    \label{fig:house107}
\end{figure}

Detailed description of the variables is available in supplementary materials. All the results for the Bayesian procedures in this section are based on the same settings and number of iterations as those used in the previous Sub-section for the \texttt{GUSTO-I} dataset. Figure \ref{fig:house107} shows the marginal PIPs for all 26 variables, along with a list of variables selected by the penalized likelihood methods.  Note that, as in previous illustrations, the penalized likelihood methods tend to select a superset of the variables selected by the Bayesian approaches, with LASSO selecting the largest superset. Furthermore, there is broad agreement among all Bayesian procedures.  For example, all techniques, except both versions of DRPEP, assign high PIP to \texttt{daysServed}.  Based on our experience with this  application, including \texttt{daysServed} in the model is sensible.  For example, length of tenure has been previously identified in the literature as an important predictor of legislator's effectiveness (e.g., see \citealp{miquel2006legislative}). One place where the different procedures do seem to disagree is whether \texttt{medianIncome} (the median income in the district represented by the legislator) explains voting behavior.  All LPEP procedures agree in providing weak to moderate evidence \textit{against} the inclusion of this variable, while most other procedures provide weak to moderate evidence \textit{in favor of it}.  Interestingly, the two versions of the DRPEP seem to disagree, with the unit information version providing weak evidence in favor of its inclusion and the hyper-g/n providing weak evidence against it.


\section{Discussion}\label{se:discussion}



The results from our theoretical and empirical studies show that the LPEP approach to creating non-informative priors for logistic regression is at least competitive, and in some cases superior, to existing techniques in terms of model selection and parameter estimation performance.  The differences are particularly striking when comparing the LPEP with the original CRPEP and DRPEP approaches proposed in \cite{fouskakis2018power}.  Aside from the increased computational expense that makes the practical application of CRPEP and DRPEP prior challenging, our empirical results show that the behavior of these procedures can be greatly affected by the choice of hyperpriors. Furthermore, both procedures can lead to unexpected results that substantially deviate from the consensus of other Bayesian and non-Bayesian methods.

We were surprised by the poor behavior of some of the LCL procedures in some of our simulation studies.  One point to note is that the setup of the simulations in Section \ref{se:simulation} ($n=500$, $p=100$) was only briefly studied in \cite{li2018mixtures}.  Indeed, most of the simulation studies in \cite{li2018mixtures} focus on settings involving fewer covariates ($p=20$), which are similar to those in the simulation study that we present in our supplementary materials.  In this lower dimensional setting, LCL behaves quite well.  Disentangling the role of the prior distribution from that of the various approximations used to speed up computation in the \texttt{BAS} package is one particular challenge in understanding these negative results. However, by comparing the results for LCL against its ``exact'' version (LCE), as well as those under the unit information ($\delta =n$) with those under the robust and hyper-g/n versions of the procedure, it would seem like the use of the Laplace approximation for the Bayes factor is not the culprit.  Instead, these results seem to driven by a combination of some sensitivity to the choice of hyperprior for $\delta$ and issues with the default approximation procedure implemented in \texttt{BAS} to integrate over $\delta$.  Interestingly, the sensitivity to the hyperprior does not seem to be present for the LPEP procedures. We believe that this is an area that is in need of further investigation in the future.

This paper focuses mostly on developing the LPEP for logistic regression.  However, the formulation is very general and can be extended to many other generalized linear models.  Many of the computational advantages of our procedure extend to binomial, negative binomial and multinomial logic models where the data augmentation approach of \cite{polson2013bayesian} can be readily applied.  This is also true for probit models in which computation can rely on the data augmentation approach of \cite{albert1993bayesian}, as well as for loglinear regression using the approach of \cite{fruhwirth2009improved}.  These extensions will be explored elsewhere.

One final brief note about our theoretical results.  Our asymptotic analyses assume that the number of covariates $p$ is fixed with the sample size $n$.  We believe that our results can be extended to situations in which $p$ grows with $n$ (as long $n$ remains larger than $p$), potentially along the lines of \cite{barber2016laplace}.  However, this requires the careful introduction of additional constraints that ensure that the maximum likelihood estimates under the training sample remains finite as both $p$ and $n$ (and therefore, as $n^{*}$ grow).  We plan to explore this question in our future work.

\section*{Additional materials}

The data sets used in the simulation studies in Section \ref{se:simulation} along with real data sets (for e.g. \texttt{Urinary}, \texttt{endometrial}, \texttt{GUSTO}, and \texttt{House107}) and code for Laplace PEP methodology discussed in the paper with options to implement LCE and \cite{bove2011hyper}'s prior  are available at \url{https://github.com/Anupreet-Porwal/LPEP}. 
Code to replicate the results in Section \ref{se:simulation} and \ref{se:realdata} is available at \url{https://github.com/Anupreet-Porwal/LPEP-Paper-Analysis}.  Supplementary materials, which include a further set of simulation studies along with the analysis of the \texttt{endometrial}  and \texttt{PIMA} data set \citep{heinze2002solution,agresti2015foundations} is  available from the authors.

\section*{Acknowledgements}

We would like to thank Dimitris Fouskakis, Ioannis Ntzourfras and Konstantinos Perrakis for sharing their code for the CRPEP and DRPEP, which we used in our analyses. 

\appendix

\section{Proof of Theorem \ref{th:existence} (Existence of MLEs)}\label{ap:MLEinsubmodels}

The two conditions in the theorem together imply that the maximum likelihood estimator for the full model, 
$$
\hat{\bbeta}_{\bgam_F} = \argmax_{\bbeta_{\bgam_F}} \, \ell_{\bgam_F}(\bbeta_{\bgam_F}) , 
$$
exists, is finite, and, furthermore, 
is unique (e.g., see \citealp{makelainen1981existence}). 

Now, note that for any other model 
$\bgam$, maximizing $\ell_{\bgam}(\bbeta_{\bgam})$ is equivalent to maximizing  $\ell_{\bgam_F}(\bbeta_{\bgam_F})$ subject to the constraint $\bbeta_{\bgam_F} \in S_{\bgam}$, where $S_{\bgam} = \{ \bftheta \in S_{\bgam_F} : \theta_j = 0 \mbox{ if } \gamma_j=0  \}$ and that $S_{\bgam}$ is also an open connected set for all $\bgam$.
Because $\ell_{\bgam}(\bbeta_{\bgam_F})$ is continuous and strongly concave, then its restriction to $S_{\bgam}$  
is also continuous and strongly concave for any $\bgam$.  Furthermore, we also have $\lim_{\bbeta_{\bgam} \to \partial S_{\bgam}}\ell_{\bgam}(\bbeta_{\bgam}) = -\infty$ because $\partial S_{\bgam} \subset \partial S_{\bgam_F}$.  Therefore, $\hat{\bbeta}_{\bgam}$ also exists and is finite and unique for any $\bgam$. \hfill $\Box$

\section{Proof of Theorem  \ref{th:tails} (Tail behavior)}\label{ap:tails}

We develop the argument only for $\pi_{\bgam}^{LPEP-R}(\bbeta_{\bgam})$.  The proof for $\pi_{\bgam}^{LPEP-HGN}(\bbeta_{\bgam})$ follows along almost identical lines.  Note that
\begin{align*}
\lim_{s \to \infty} \frac{\zeta^{R}(s \mid \bv, \bgam)}{ \left( 1 + s^2/(p_{\bgam}+1) \right)^{-\frac{p_{\bgam}+2}{2}}} = 
\sum_{\by^{*}} m^{*}(\bfy^{*}) \lim_{s \to \infty}
 \frac{\zeta^{R}(s \mid \bv, \bgam, \by^{*})}{ \left( 1 + s^2/(p_{\bgam}+1) \right)^{-\frac{p_{\bgam}+2}{2}}}  ,
\end{align*}
where
\begin{multline*}
\zeta^{R}(s \mid \bv, \bgam, y^{*}) = \left. \pi_{\bgam}^{LPEP-R}(\bbeta_{\bgam} \mid \by^{*})  \right|_{\bbeta_{\bgam} = s \bv} \\
= \int_{0}^{\infty} \left( \frac{1}{2\pi\delta} \right)^{\frac{p_{\bgam}+1}{2}}
\left| \bH_{\bgam}(\bfy^{*}) \right|^{1/2} \exp\left\{ -\frac{1}{2\delta} \left(s \bv - \hat{\bbeta}_{\bgam}(\by^*) \right)^T \bH_{\bgam}(\bfy^{*})
\left(s \bv - \hat{\bbeta}_{\bgam}(\by^*) \right)\right\} f^R(\delta|\bgam) d\delta 
\end{multline*}
is conditional on a given training sample $\by^{*}$.  (We can exchange the summation and the limit in this case because, for any $n^{*}$, the number of potential training samples is finite.) In the sequel, it will also be important to remember that $m^{*}(\by^{*})$ is defined so that the maximum likelihood estimators exist for any sample $\by^{*}$.

To simplify notation, define
$$
\zeta^{R-C}(s \mid \bv, \bgam, y^{*}) = 
\int_{0}^{\infty} \left( \frac{1}{2\pi\delta} \right)^{\frac{p_{\bgam}+1}{2}}
\left| \bH_{\bgam}(\bfy^{*}) \right|^{1/2} \exp\left\{ -\frac{1}{2\delta} s^2 \bv^T \bH_{\bgam}(\bfy^{*}) \bv
\right\} f^R(\delta|\bgam) d\delta .
$$
Note that
\begin{multline*}
\lim_{s \to \infty} \frac{\zeta^{R}(s \mid \bv, \bgam, \by^{*})}{ \left( 1 + s^2/(p_{\bgam}+1) \right)^{-\frac{p_{\bgam}+2}{2}}}
= \\
\lim_{s \to \infty} \frac{\zeta^{R}(s \mid \bv, \bgam, \by^{*})}{\zeta^{R-C}(s \mid \bv, \bgam, y^{*})} \times
\lim_{s \to \infty} \frac{ \left(s^2 \right)^{-\frac{p_{\bgam}+2}{2}}}{\left( 1 + s^2/(p_{\bgam}+1) \right)^{-\frac{p_{\bgam}+2}{2}}}  \times
\lim_{s \to \infty} \frac{\zeta^{R-C}(s \mid \bv, \bgam, y^{*})}{\left(s^2 \right)^{-\frac{p_{\bgam}+2}{2}}} .
\end{multline*}

Clearly, the first two limits converge to finite functions that $\bv$, the training sample $\by^{*}$ and/or the model $\bgam$.  Hence,
\begin{align*}
\lim_{s \to \infty} \frac{\zeta^{R}(s \mid \bv, \bgam, \by^{*})}{ \left( 1 + s^2/(p_{\bgam}+1) \right)^{-\frac{p_{\bgam}+2}{2}}}
& = c_{\bgam,1}(\bv, y^{*}) 
\lim_{s \to \infty} \left(s^2 \right)^{\frac{p_{\bgam}+2}{2}} \zeta^{R-C}(s \mid \bv, \bgam, y^{*}) ,
\end{align*}
where $0 < c_{\bgam,1}(\bv, y^{*}) < \infty$. Plugging in $\zeta^{R-C}(s \mid \bv, \bgam, y^{*})$ and $f^R(\delta|\bgam)$, we can write 
\begin{multline*}
    \lim_{s \to \infty} \frac{\zeta^{R}(s \mid \bv, \bgam, \by^{*})}{ \left( 1 + s^2/(p_{\bgam}+1) \right)^{-\frac{p_{\bgam}+2}{2}}}
= 
c_{\bgam,1}(\bv, y^{*})\lim_{s \to \infty} \left(s^2 \right)^{\frac{p_{\bgam}+2}{2}}\times \\
\int_{0}^{\infty} \left( \frac{1}{2\pi\delta} \right)^{\frac{p_{\bgam}+1}{2}}
\left| \bH_{\bgam}(\bfy^{*}) \right|^{1/2} \exp\left\{ -\frac{1}{2\delta} s^2 \bv^T \bH_{\bgam}(\bfy^{*}) \bv
\right\}\times \\
\frac{1}{2 (p_{\bgam}+1)^{1/2}}\frac{ (n^{*}+1)^{1/2}}{(\delta+1)^{3/2}}\mathbf{1} \left( \delta>\frac{n^{*} - p_{\bgam}}{p_{\bgam}+1} \right) d\delta .
\end{multline*}

Substituting $\lambda=\left(\frac{n^*+1 }{p_{\bgam}+1}\right)\frac{1}{\delta+1}$, we can write above equation as 
\begin{multline*}
     \lim_{s \to \infty} \frac{\zeta^{R}(s \mid \bv, \bgam, \by^{*})}{ \left( 1 + s^2/(p_{\bgam}+1) \right)^{-\frac{p_{\bgam}+2}{2}}}
= 
c_{\bgam,1}(\bv, y^{*})\left( \frac{1}{2\pi} \right)^{\frac{p_{\bgam}+1}{2}} \left| \bH_{\bgam}(\bfy^{*}) \right|^{1/2}\times \\ 
\lim_{s \to \infty}  \left(s^2 \right)^{\frac{p_{\bgam}+2}{2}} \int_0^1 \left(\frac{\lambda}{m-\lambda}\right)^{\frac{p_{\bgam}+1}{2}}\lambda^{-\frac{1}{2}}\exp\left\{-\left(\frac{\lambda}{m-\lambda}\right)qs^2 \right\} d\lambda
\end{multline*}
where $q=\frac{1}{2}\bv^T \bH_{\bgam}(\bfy^{*}) \bv$, $m=\frac{n^*+1}{p_{\bgam}+1}>1$  since $n^*>p_{\bgam}$. Now, from Lemma 2 in \cite{bayarri2012criteria}, 
$$
\lim_{s \to \infty}  \left(s^2 \right)^{\frac{p_{\bgam}+2}{2}} \int_0^1 \left(\frac{\lambda}{m-\lambda}\right)^{\frac{p_{\bgam}+1}{2}}\lambda^{-\frac{1}{2}}\exp\left\{-\left(\frac{\lambda}{m-\lambda}\right)qs^2 \right\} d\lambda = c_{\bgam,2}(\bv, y^{*}) ,
$$

where $0 < c_{\bgam,2}(\bv, y^{*}) < \infty$, and therefore
$$
\lim_{s \to \infty} \frac{\zeta^{R}(s \mid \bv, \bgam, \by^{*})}{ \left( 1 + s^2/(p_{\bgam}+1) \right)^{-\frac{p_{\bgam}+2}{2}}}
= c_{\bgam,1}(\bv, y^{*}) c_{\bgam,2}(\bv, y^{*}) = c_{\bgam,3}(\bv, y^{*}) .
$$

To complete the proof simply define $c_{\bgam}(\bv) = \sum_{\by^{*}} m^{*}(\bfy^{*}) c_{\bgam,3}(\bv, y^{*}) $.  Since we have a finite number of terms in the sum and each term is both positive and finite, so is $c_{\bgam}(\bv)$.

\section{Proof of Theorem  \ref{th:intrinsicpriorconsistency} (Intrinsic consistency)}\label{ap:properties2}

Note that if $\by^{*}$ leads to separability, then $\by^{**} = \mathbf{1} - \by^{*}$ does as well.  Hence, our choice of $m^{*}\left(\by^{*} \mid \bX \right)$ is symmetric and, as $n^{*} \to \infty$, $\hat{\bbeta}_{\bgam}(\by^{*}) \overset{p}{\to} \mathbf{0}$ under $m^{*}$. In turn, this implies that $\hat{\theta}_{\bgam,i}(\by^{*}) \overset{p}{\to} 1/2$ for all $i$, and therefore $\frac{1}{n}\bX_{\bgam}' \bW_{\bgam}(\by^{*}) \bX_{\bgam} \overset{p}{\to} \frac{1}{4} \bfSigma_{\bgam}$. This completes the proof when $g=n$.

In the case where $\delta$ is given a prior distribution define $\delta = n^{*} \delta^{*}$.  Then, under $f^{HGn}(\delta)$, $\delta^{*}$ has density
$$
f^{HGn}(\delta^{*}) = \left( 1 + \delta^{*}\right)^{-2} 
$$
which is a proper, non-degenerate prior.  The argument for the robust prior follows along similar lines.

\hfill $\Box$

\section{Proof of Theorem \ref{th:asympconsistency} (Model selection consistency)}\label{ap:properties1}

Since 
$$
Pr(\bgam = \bgam_{T} \mid \by)=\frac{1}{1 + \sum_{\bgam \ne \bgam_T}\frac{f(\bgam)}{f(\bgam_T)} \frac{m^{LPEP}_{\bgam}(\by)}{m^{LPEP}_{\bgam_T}(\by)}}
$$
it is enough to show that 
$$
\frac{m^{LPEP}_{\bgam}(\by)}{m^{LPEP}_{\bgam_T}(\by)}  \xrightarrow[n \to \infty]{P} 
  0 .
$$ 

The proof follows along similar lines as that in \cite{li2018mixtures}.  We start by assuming similar regularity conditions:
\begin{itemize}
    \item[(i)] The true model $\bgam_T$ is among the $2^p$ models under consideration, with $p$ fixed.
    \item[(ii)] For every $i=1,2,\ldots$, the vector $\bx_i$ is such that $\left\| \bx_i \right\|_2$ is bounded by a constant.
    \item[(iii)] For all $n$, the smallest eigenvalue of $\frac{1}{n}\bX^T\bX$ is bounded from below by a positive constant.
\end{itemize}
Note that these conditions imply the weak consistency and asymptotic normality of the maximum likelihood estimators for all models under consideration (e.g., see \citealp{fahrmeir1985consistency}).

Consider first 
$m_{\bgam}(\by \mid \by^{*}) = \int f_{\bgam} (\by \mid \betag)  \phi_{p_{\bgam}+1} \left( \betag \mid \hat{\bbeta}_{\bgam}\left( \by^{*}\right), n \bH_{\bgam}^{-1}\left( \by^{*}\right) \right) d \betag$.  Using a Laplace approximation of $f_{\bgam} (\by \mid \betag)$,
\begin{align*}
    f_{\bgam} (\by \mid \betag) &= 
    f_{\bgam} \left( \by \mid \hat{\bbeta}_{\bgam} \right)
    \exp \left\{ - \frac{1}{2}
    \left( \betag - \hat{\bbeta}_{\bgam} \right)^{T}
    \bH_{\bgam}
    \left( \betag - \hat{\bbeta}_{\bgam} \right)
    \right\}
    \exp\left\{ R\left(\betag , \hat{\bbeta}_{\bgam}\right) \right\}
\end{align*}
where $R\left(\betag , \hat{\bbeta}^{*}_{\bgam}\right)$ is the residual form the second order Taylor expansion of the loglikelihood.  Note that, to simplify notation, we have let $\hat{\bbeta}_{\bgam} = \hat{\bbeta}_{\bgam} \left( \by \right)$ and $\bH_{\bgam} = \bH_{\bgam}\left( \by \right)$.

Hence, following \citep{tierney1986accurate},
\begin{multline*}
    m_{\bgam}\left( \by \mid \by^{*} \right) = f_{\bgam} \left( \by \mid \hat{\bbeta}_{\bgam} \right)
    n^{-\frac{p_{\bgam}+1}{2}} \left( \frac{\left| \bH_{\bgam} \right|}{\left| \bE_{\bgam}) \right|}\right)^{1/2} \\
    \exp\left\{ - \frac{1}{2} \left[ \hat{\bbeta}_{\bgam}^T \bH_{\bgam} \hat{\bbeta}_{\bgam} + \frac{1}{n} \hat{\bbeta}^{*T}_{\bgam} \bH^{*}_{\bgam} \hat{\bbeta}^{*}_{\bgam} - 
    \bd^T_{\bgam} \bE_{\bgam}^{-1}\bd_{\bgam} \right]
    \right\}
    \left( 1 + \mathcal{O}  \left(\frac{1}{n}\right)\right)
\end{multline*}
where
\begin{align*}
    \bd_{\bgam} &= \bH_{\bgam}\hat{\bbeta}_{\bgam} + \frac{1}{n} \bH^{*}_{\bgam}\hat{\bbeta}_{\bgam}^{*} , &
    \bE_{\bgam} &= \bH_{\bgam} + \frac{1}{n} . \bH^{*}_{\bgam} 
\end{align*}
As before, we simplify notation by letting $\hat{\bbeta}_{\bgam}^{*} = \hat{\bbeta}_{\bgam} \left( \by^{*} \right)$,  and $\bH^{*}_{\bgam} = \bH_{\bgam}\left( \by^{*} \right)$.  From this, 
\begin{multline*}
    \frac{m^{LPEP}_{\bgam}(\by)}{m^{LPEP}_{\bgam_T}(\by)} = \frac{f_{\bgam} \left( \by \mid \hat{\bbeta}_{\bgam} \right)}
    {f_{\bgam_T} \left( \by \mid \hat{\bbeta}_{\bgam_T} \right)} n^{(p_{\bgam_T} - p_{\bgam})/2} \\
    \frac{\mathsf{E}_{y^{*}} \left\{  \left( \frac{\left| \bH_{\bgam} \right|}{\left| \bE_{\bgam}) \right|} 
    \right)^{1/2} \exp\left\{ - \frac{1}{2} \left[ \hat{\bbeta}_{\bgam}^T \bH_{\bgam} \hat{\bbeta}_{\bgam} + \frac{1}{n} \hat{\bbeta}^{*T}_{\bgam} \bH^{*}_{\bgam} \hat{\bbeta}^{*}_{\bgam} - 
    \bd^T_{\bgam} \bE^{-1}_{\bgam}\bd_{\bgam} \right]
    \right\}\right\}}
    {\mathsf{E}_{y^{*}} \left\{ \left( \frac{\left| \bH_{\bgam_T} \right|}{\left| \bE_{\bgam_T}) \right|} 
    \right)^{1/2} \exp\left\{ - \frac{1}{2} \left[ \hat{\bbeta}_{\bgam_T}^T \bH_{\bgam_T} \hat{\bbeta}_{\bgam_T} + \frac{1}{n} \hat{\bbeta}^{*T}_{\bgam_T} \bH^{*}_{\bgam_T} \hat{\bbeta}^{*}_{\bgam_ T} - 
    \bd^T_{\bgam_T} \bE^{-1}_{\bgam_T}\bd_{\bgam_T} \right]
    \right\} \right\}}
    \, \mathcal{O}(1)
\end{multline*}

Consider first the term
$$
\frac{f_{\bgam} \left( \by \mid \hat{\bbeta}_{\bgam} \right)}
    {f_{\bgam_T} \left( \by \mid \hat{\bbeta}_{\bgam_T} \right)} n^{(p_{\bgam_T} - p_{\bgam})/2} .
$$
This is just Schwartz criterion, which is well known to be consistent in this setting.  Indeed, under the regularity conditions (i)-(iii), Lemma A.3 in the supplementary materials of \cite{li2018mixtures} can be applied to show that, as $n$ increases,  the likelihood ratio $\Lambda_{\bgam,\bgam_T} = \frac{f_{\bgam} \left( \by \mid \hat{\bbeta}_{\bgam} \right)}
{f_{\bgam_T} \left( \by \mid \hat{\bbeta}_{\bgam_T} \right)}$ has the following behavior.
\begin{itemize}
    \item[(a)] If $\bgam_T \subset \bgam$, then $\Lambda_{\bgam,\bgam_T} = \mathcal{O}_P(1)$. 
    \item[(b)] If $\bgam_T \not\subset \bgam$ then $\Lambda_{\bgam,\bgam_T} = \mathcal{O}_P\left(e^{-c_{\bgam}n}\right)$ for some positive constant $c_{\bgam}$.
\end{itemize}
Hence, if $\bgam_T \subset \bgam$ then necessarily $p_{\bgam_T} < p_{\bgam}$ and
$$
\frac{f_{\bgam} \left( \by \mid \hat{\bbeta}_{\bgam} \right)}
    {f_{\bgam_T} \left( \by \mid \hat{\bbeta}_{\bgam_T} \right)} n^{(p_{\bgam_T} - p_{\bgam})/2} = \mathcal{O}_P(n^{-c})
$$
for some positive constant $c$ and therefore it tends to zero as $n\to \infty$.  On other hand, $\bgam_T \not\subset \bgam$
$$
\frac{f_{\bgam} \left( \by \mid \hat{\bbeta}_{\bgam} \right)}
    {f_{\bgam_T} \left( \by \mid \hat{\bbeta}_{\bgam_T} \right)} n^{(p_{\bgam_T} - p_{\bgam})/2} = 
    \mathcal{O}_P\left(e^{-c_{\bgam}n}\right)
$$
for some positive constant $c_{\bgam}$ no matter whether $p_{\bgam_T} < p_{\bgam}$ or $p_{\bgam_T} \ge p_{\bgam}$, and again it converges to zero as $n\to \infty$.

All that remains now is to show that 
\begin{equation}\label{eq:BIChigherorderresidual}
 \frac{\mathsf{E}_{y^{*}} \left\{  \left( \frac{\left| \bH_{\bgam} \right|}{\left| \bE_{\bgam}) \right|} 
    \right)^{1/2} \exp\left\{ - \frac{1}{2} \left[ \hat{\bbeta}_{\bgam}^T \bH_{\bgam} \hat{\bbeta}_{\bgam} + \frac{1}{n} \hat{\bbeta}^{*T}_{\bgam} \bH^{*}_{\bgam} \hat{\bbeta}^{*}_{\bgam} - 
    \bd^T_{\bgam} \bE^{-1}_{\bgam}\bd_{\bgam} \right]
    \right\}\right\}}
    {\mathsf{E}_{y^{*}} \left\{ \left( \frac{\left| \bH_{\bgam_T} \right|}{\left| \bE_{\bgam_T}) \right|} 
    \right)^{1/2} \exp\left\{ - \frac{1}{2} \left[ \hat{\bbeta}_{\bgam_T}^T \bH_{\bgam_T} \hat{\bbeta}_{\bgam_T} + \frac{1}{n} \hat{\bbeta}^{*T}_{\bgam_T} \bH^{*}_{\bgam_T} \hat{\bbeta}^{*}_{\bgam_ T} - 
    \bd^T_{\bgam_T} \bE^{-1}_{\bgam_T}\bd_{\bgam_T} \right]
    \right\} \right\}}    
\end{equation}
is, at most,  $\mathcal{O}_P(1)$.  To do this, we consider the behavior of 
$$
\left( \frac{\left| \bH_{\bgam} \right|}{\left| \bE_{\bgam}) \right|} 
    \right)^{1/2} \exp\left\{ - \frac{1}{2} \left[ \frac{1}{n} \hat{\bbeta}^{*T}_{\bgam} \bH^{*}_{\bgam} \hat{\bbeta}^{*}_{\bgam} + \hat{\bbeta}_{\bgam}^T \bH_{\bgam} \hat{\bbeta}_{\bgam} - 
    \bd^T_{\bgam} \bE^{-1}_{\bgam}\bd_{\bgam} \right]
    \right\}    
$$
in probability as both $\by$ and $\by^{*}$ grow.  Because of dominated convergence, the behavior in probability with respect to  $\by^{*}$ is the same as that in expectation.

\begin{itemize}
    \item First, note that $\frac{\left| \bH_{\bgam} \right|}{\left| \bE_{\bgam} \right|} = \left| \mathbf{I} + \frac{1}{n} \bH^{-1}_{\bgam}\bH^{*}_{\bgam} \right|^{-1} = \left| \mathbf{I} + \frac{1}{n} \left( n \bH^{-1}_{\bgam} \right) \left( \frac{1}{n} \bH^{*}_{\bgam} \right) \right|^{-1}$.  Under regularity conditions (ii) and (iii), $\frac{1}{n} \bH_{\bgam} \xrightarrow[n \to \infty]{P} \bJ_{\bgam}$ and $\frac{1}{n} \bH^{*}_{\bgam} \xrightarrow[n \to \infty]{P} \bJ^{*}_{\bgam}$ where $\bJ_{\bgam}$ and $\bJ^{*}_{\bgam}$ are $(p_{\bgam}+1) \times (p_{\bgam}+1)$ constant and strictly positive definite matrices with finite entries (please see below).  Hence,
    $$
    \left| \mathbf{I} + \frac{1}{n} \bH^{-1}_{\bgam}  \bH^{*}_{\bgam}  \right|^{-1} \xrightarrow[n \to \infty]{P} 1.
    $$
    
    We only elaborate on the proof that $\frac{1}{n} \bH_{\bgam} \xrightarrow[n \to \infty]{P} \bJ_{\bgam}$; the argument for $\frac{1}{n} \bH^{*}_{\bgam}$ is analogous.  Furthermore, we focus on the  diagonal elements of the matrix since the off-diagonal elements are bounded by the diagonal ones because the matrix is, by construction at least semi-positive definite. Since $\left\| \bx_i \right\|_2 < c$ for all $i$ and some constant $c$, we have $| x_{i,j} | < c $.  Therefore $\left[ \frac{1}{n} \bH_{\bgam} \right]_{j,j} = \frac{1}{n} \sum_{i=1}^{n} \hat{w}_{\bgam,i} x_{i,j}^2 < \frac{c^2}{n} \sum_{i=1}^{n} \frac{\left( 1 + \exp\left\{ -\bx_i^{T} \hat{\bbeta}_{\bgam}\right\} \right)^2}{\exp\left\{ -\bx_i^{T} \hat{\bbeta}_{\bgam} \right\}}$.  Now, the maximum likelihood estimator $\hat{\bbeta}_{\bgam}$ is not necessarily consistent when $\bgam \ne \bgam_T$, but it does converge in probability to a finite constant (e.g., see \citealp{fahrmeir1990maximum}).  Combined with the fact that the $\bx_i$s have compact support, this implies that $\left(1 + \exp\left\{ - \bx_i^T\hat{\bbeta}_{\bgam} \right\} \right)^{-1}$ converges in probability to a constant $\theta^{(0)}_{\bgam,i}$ that, for all $i$, is bounded away from 0 and 1, i.e., $0 < a \le \theta^{(0)}_{\bgam,i} \le b < 1$ for some constants $a$ and $b$.  Therefore, $\left[ \frac{1}{n} \bH_{\bgam} \right]_{j,j}   <  \frac{c^2}{\min\{ a(1-a),b(1-b) \}} < \infty$.  

    \item Next, consider  $\frac{1}{n}\hat{\bbeta}^{*T}_{\bgam} \bH^{*}_{\bgam} \hat{\bbeta}^{*}_{\bgam}$.  Because of our choice for $m^{*}(\by^{*})$, the results in \cite{fahrmeir1990maximum} imply that  $\frac{1}{n}\hat{\bbeta}^{*T}_{\bgam} \bH^{*}_{\bgam} \hat{\bbeta}^{*}_{\bgam}$ converges in distribution to a point mass at zero.  This is because the null model is the closest model (in the Kullback–Leibler sense) to $m^{*}(\by^{*})$. Therefore we also have
    $$
    \frac{1}{n}\hat{\bbeta}^{*T}_{\bgam} \bH^{*}_{\bgam} \hat{\bbeta}^{*}_{\bgam} \xrightarrow[n \to \infty]{P} 0 .
    $$
    

    \item Finally, consider $\hat{\bbeta}^{T}_{\bgam} \bH_{\bgam} \hat{\bbeta}_{\bgam} - 
    \bd^T_{\bgam} \bE^{-1}
    _{\bgam}\bd_{\bgam}$.  Substituting back the expressions for $\bd_{\bgam}$ and $\bE_{\bgam}$, expanding the quadratic form and combining similar terms we have:
    \begin{multline*}
      \hat{\bbeta}^{T}_{\bgam} \bH_{\bgam} \hat{\bbeta}_{\bgam} - 
    \bd^T_{\bgam} \bE^{-1}
    _{\bgam}\bd_{\bgam} =  \hat{\bbeta}^{T}_{\bgam} \left[ \bH_{\bgam} - \bH_{\bgam}\left(\bH_{\bgam} + \frac{1}{n} \bH_{\bgam}^{*}\right)^{-1}\bH_{\bgam}  \right] \hat{\bbeta}_{\bgam} \\
    + \frac{2}{n}\hat{\bbeta}^{T}_{\bgam} \bH_{\bgam}\left(\bH_{\bgam} + \frac{1}{n} \bH_{\bgam}^{*}\right)^{-1}\bH^{*}_{\bgam} \hat{\bbeta}^{*}_{\bgam}
    - \frac{1}{n^2}\hat{\bbeta}^{*T}_{\bgam} \bH^{*}_{\bgam}\left(\bH_{\bgam} + \frac{1}{n} \bH_{\bgam}^{*}\right)^{-1}\bH^{*}_{\bgam} \hat{\bbeta}^{*}_{\bgam}
    \end{multline*}
    Now, from the Woodbury matrix identity we have
    $$
    \left[ \bH_{\bgam} - \bH_{\bgam}\left(\bH_{\bgam} + \frac{1}{n} \bH_{\bgam}^{*}\right)^{-1}\bH_{\bgam}  \right] = \left(\bH^{-1}_{\bgam} + n \bH_{\bgam}^{*-1}\right)^{-1}
    $$
    and therefore
    \begin{multline*}
    \hat{\bbeta}^{T}_{\bgam} \bH_{\bgam} \hat{\bbeta}_{\bgam} - 
    \bd^T_{\bgam} \bE^{-1}
    _{\bgam}\bd_{\bgam} =  \hat{\bbeta}^{T}_{\bgam} \left(\bH^{-1}_{\bgam} + n \bH_{\bgam}^{*-1}\right)^{-1}\hat{\bbeta}_{\bgam} \\
    + \frac{2}{n}\hat{\bbeta}^{T}_{\bgam} \bH_{\bgam}\left(\bH_{\bgam} + \frac{1}{n} \bH_{\bgam}^{*}\right)^{-1}\bH^{*}_{\bgam} \hat{\bbeta}^{*}_{\bgam}
    - \frac{1}{n^2}\hat{\bbeta}^{*T}_{\bgam} \bH^{*}_{\bgam}\left(\bH_{\bgam} + \frac{1}{n} \bH_{\bgam}^{*}\right)^{-1}\bH^{*}_{\bgam} \hat{\bbeta}^{*}_{\bgam}
    \end{multline*}
    
    Note that the last two terms converge in probability to $0$, while the first term is $\mathcal{O}_P\left( 1 \right)$.
    
\end{itemize}

\section{Details of the Markov chain Monte Carlo algorithm for logistic regression}\label{ap:MCMCMdetails}

Using the hierarchical representation of the LPEP prior discussed at the start of Section \ref{se:MCMC}, the posterior distribution for the augmented model can be written as
\begin{align*}
     \pi(\bgam, \betag, \bld{\omega}, \bld{y}^*, \delta \mid \bld{y}, \bld{X})\propto f_{\bgam}(\bld{y} \mid \betag,\bld{\omega})\pi_{\bgam}^{LPEP}(\betag \mid \delta, \boldsymbol{y}^*)f(\delta|\bgam)f(\bgam)f(\bld{\omega})m^*(\bld{y}^{*}) .
\end{align*}
 
From this, it is easy to devise samplers for the full conditional posterior distributions of various blocks of parameters.  We focus below on the more general setting where $\delta$ has been assigned a hyperprior.  The simplifications for the case where $\delta$ is fixed are straightforward and we do not discuss them explicitly.
\begin{enumerate}
   \item Since the prior support of $\delta$ maybe dependent on model indicator $\bgam$, the parameters $(\bgam,\delta, \betag)$ are updated jointly by sampling from $f(\bgam,\delta,  \betag \mid \by^{*},  \bfomega)$ given in  \eqref{eq:fullcondibetagam}. To do this, we write 
   $$
   f(\bgam, \delta, \betag \mid  \by^{*},  \bfomega) = f(\bgam, \delta  \mid 
   \by^{*},  \bfomega) f( \betag \mid \bgam, \delta, \by^{*},  \bfomega) , 
   $$
   where $f(\bgam,\delta \mid  \by^{*},  \bfomega) \propto \int f(\bgam,\delta, \betag \mid  \by^{*},  \bfomega) d \betag$.  

    The expression for the conditional posterior $f(\bgam,\delta \mid  \by^{*}, \bfomega)$, up to a proportionality constant, is given by \eqref{eq:marginallikelihood}. 
    To generate samples from it, we generate proposal by combining a random walk Metropolis-Hastings algorithm for $\bgam$ \citep{george1997approaches} and a reflective Gaussian random walk for $\delta$ (similar to section 2.1 of \citealp{thawornwattana2018designing}). More specifically, we factorize the joint proposal for  $(\bgam,\delta)$ as: 
     \begin{align*}
         q\left(\delta^{(prop)}, {\bgam}^{(prop)} \mid \delta,\bgam \right) = q\left({\bgam}^{(prop)} \mid \bgam \right)q\left(\delta^{(prop)}\mid \delta, {\bgam}^{(prop)}\right).
     \end{align*}
    For $q({\bgam}^{(prop)} \mid \bgam)$, we use a symmetric random walk proposal similar to equation (46) of \cite{george1997approaches} as follows:
 \begin{itemize}
     \item We define two probability vectors $p_1=(0.9,0.1)$ and $p_2=(0.6,0.2,0.15,0.05)$.
     \item Each time, we decide on one of two types of moves according to the probability vector $p_1$.
     \begin{itemize}
         \item If a move type 1 is selected, then the proposed new model $\bgam^{(prop)}$ is generated by randomly flipping $d\in \{1,2,3,4\}$ components of $\bgam$ with probability $p_{2,d}$.  The components of $\bgam$ to be flipped are selected uniformly at random given $d$.
         \item If a move type 2 is selected, then the proposed model $\bgam^{(prop)}$ is generated by removing one variable currently included in the model and replacing it with a variable that is currently excluded, leaving the dimensionality of the model unchanged.  The variables to add and remove are chosen uniformly at random within each set.
     \end{itemize}
 \end{itemize}
 
 Next, given $\bgam^{(prop)}$, we propose $\delta$ using a reflective Gaussian random walk with a left reflection boundary $a_{\bgam^{(prop)}}$. More specifically, we define $\delta^{(prop)}=a_{\bgam^{(prop)}} + |\epsilon -a_{\bgam^{(prop)}}|$ where $\epsilon\sim N(\delta, \tau^2)$ and $a_{\bgam}=0$ under the hyper-g/n prior and
 $a_{\bgam}=\frac{n-p_{\bgam}}{p_{\bgam}+1}$ under the robust prior. In both cases we found $\tau= n/2$ to be an efficient tuning parameter in our studies.
 
 Since the proposal distribution of $\bgam$, given by $q({\bgam}^{(prop)} \mid \bgam)$ is symmetric, the proposed model $(\bgam^{(prop)},\delta^{(prop)})$ is then accepted with probability
 $$
 \min\left\{\frac{f(\bgam^{(prop)},\delta^{(prop)} \mid \by^{*},  \bfomega) }{f(\bgam, \delta \mid \by^{*},  \bfomega) }\frac{q(\delta\mid \delta^{(prop)}, {\bgam})}{q(\delta^{(prop)}\mid \delta, {\bgam}^{(prop)})},1\right\},
 $$ 
 where
 \begin{multline*}
 q\left( \delta^{(prop)} \mid \delta, {\bgam}^{(prop)} \right) =
 \frac{1}{\sqrt{2 \pi}\tau} \Bigg[ 
\exp\Bigg\{ -\frac{1}{2 \tau^2} (\delta^{(prop)} - \delta)^2\Bigg\} + \\
  \exp\Bigg\{ -\frac{1}{2 \tau^2} (2a_{{\bgam}^{(prop)}} -\delta^{(prop)} - \delta)^2\Bigg\} 
  \Bigg], \quad\quad \delta^{(prop)}\geq a_{\bgam^{(prop)}}.
\end{multline*}


Note that,  when move type 2 is selected, since $p_{\bgam}$ remains unchanged, $q(\delta\mid \delta^{(prop)}, {\bgam})=q(\delta^{(prop)}\mid \delta, {\bgam}^{(prop)})$ and the acceptance probability simplifies further. Similar simplification is observed under hyper-g/n prior since proposal reflection boundary, $a_{\bgam}=0$. 

Once the model $(\bgam,\delta)$ has been sampled, the regression coefficients can be updated using the fact that $\betag \mid \bgam, \delta, \by^{*}, \bfomega \sim \normal \left( \bm_{\bfomega}, \bV_{\bfomega}\right)$,
where $\bm_{\bfomega}$ and $\bV_{\bfomega}$ are given in \eqref{eq:parambetapost}.

\item  A posteriori, the entries of $\bfomega$ are conditionally independent from each other.  Following \cite{polson2013bayesian}, it is straightforward to see that
$ \omega_i \mid \bgam, \betag, \delta, \by^{*} \sim PG(1,\bld{x}_{i,\bgam}^T\betag)$.  Implementations of the samplers for the P\`olya-Gamma distribution are available, for example, in the \texttt{R} package  \texttt{BayesLogit}.

\item The conditional distribution of $\bld{y}^*$ is proportional to 
\begin{align*}
    \pi(\bld{y}^* \mid \delta,\betag)\propto \pi_{\bgam}^{LPEP}(\betag \mid \delta, \boldsymbol{y}^*)m^*(\bld{y}^*)
\end{align*}

While this distribution is supported over the finite set $\{ 0,1 \}^{n}$, a direct sampler is difficult to construct in part because of its (typically) large size of the support.  Hence, we rely again on Metropolis-Hastings steps.

In order to ensure adequate mixing of the algorithm, we consider both  local and global proposals.  At each iteration, the algorithm selects local moves with probability $0.7$ and global moves with probability $0.3$.
\begin{itemize}
    \item For the local moves, we  propose new  $\by^{*(prop)}$ by randomly flipping $d\in \{1,2,3,4,5\}$ components of $\by^{*}$ with probability $(0.5,0.2,0.15,0.10,0.05)$.  The components of $\by^{*}$ to be flipped are selected uniformly at random given $d$. Because this proposal is symmetric, the acceptance probability for this move is simply
    $$
    \min\left\{1, \frac{\pi_{\bgam}^{LPEP}(\betag \mid \delta, \boldsymbol{y}^{*(prop)})m^*(\boldsymbol{y}^{*(prop)})}{\pi_{\bgam}^{LPEP}(\betag \mid \delta, \boldsymbol{y}^*)m^*(\bld{y}^*)}\right\}
    $$
    
    \item For the global moves, we use an independent proposal similar to that used by \cite{fouskakis2018power}, 
    $q(\bld{y}^*) =
    \prod_{i=1}^n \text{Berl}( {\pi_i}^{1/\delta})$
    where
    \begin{align*}
    \pi_i^* &=\frac{\pi_0^{1/n}\pi_{i,\bgam_{-1}}^{1/\delta}}{\pi_0^{1/n}\pi_{i,\bgam_{-1}}^{1/\delta}+(1-\pi_0)^{1/n}(1-\pi_{i,\bgam_{-1}})^{1/\delta}},
    \end{align*}
    $\pi_0=\frac{1}{1+\exp{(-\beta_0)}}$,   $\pi_{i,\bgam_{-1}}^*=\frac{1}{1+\exp{(-\bld{x}_{i,{\bgam}_{-1}}^T\bld{\beta}_{\bgam_{-1}}})}$, and $\bld{\beta}_{\bgam_{-1}}$ and $\bgam_{-1}$ represent the coefficient vector and indicator variable excluding the intercept term.  The associated acceptance probability is then
    \begin{align*}
    \min\left\{1, \frac{\pi_{\bgam}^{PEP}\left(\betag \mid \delta, {\boldsymbol{y}^{*(prop)}}\right)m^{*}\left({\boldsymbol{y}^{*(prop)}}\right)q(\boldsymbol{y}^*)}{\pi_{\bgam}^{PEP}(\betag \mid \delta, \boldsymbol{y}^*)m^*(\bld{y}^*)q\left({\boldsymbol{y}^{*(prop)}}\right)}\right\}
    \end{align*}
\end{itemize}

\item The fact that $\delta$ is jointly sampled with $\bgam$ in step 1 above means that the algorithm might be slow to mix.  In order to address this issue, we incorporate an additional sampler for $\delta$ alone. The target full conditional distribution is given by: 
\begin{align*}
    \pi(\delta \mid \betag,\bld{y}^*,\bgam)\propto \pi_{\bgam}^{PEP}(\betag \mid \delta, \boldsymbol{y}^*)f(\delta|\bgam)
\end{align*}

For the prior distributions we discuss in this paper, this full posterior conditional distribution does not belong to a known family. Hence, we again use a Metropolis-Hastings algorithm to sample $\delta$ that mimics what we did in step 1.  In particular, we propose new values for $\delta$ from a reflective Gaussian distribution centered around the current value of $\delta$ and with scale $\tau=n/2$ and a left reflective boundary $a_{\bgam}=0$ fr the hyper-g/n prior and  $a_{\bgam}=\frac{n-p_{\bgam}}{p_{\bgam}+1}$ for the robust prior.  The proposed values are then accepted with probability:
\begin{align*}
    \min\left\{ 1, \frac{\pi_{\bgam}^{LPEP}\left(\betag \mid \delta^{(prop)}, \boldsymbol{y}^*\right)p\left(\delta^{(prop)}\right)  }{\pi_{\bgam}^{LPEP}(\betag \mid \delta, \boldsymbol{y}^*)p(\delta)} \frac{q(\delta\mid \delta^{(prop)}, {\bgam})}{q(\delta^{(prop)}\mid \delta, {\bgam})} \right\}.
\end{align*}

\end{enumerate}

\clearpage

\bibliographystyle{bka}
\bibliography{LPEP}

\end{document}


\renewcommand{\thetable}{S\arabic{table}}
\setcounter{figure}{0}

\renewcommand{\thefigure}{S\arabic{figure}}
\begin{center}
{ \LARGE Supporting Information for ``Laplace Power-expected-posterior priors for generalized linear models with applications to logistic regression"} \\
\vspace{3mm}
{ \large Anupreet Porwal and Abel Rodriguez}\\
 Department of Statistics, University of Washington Seattle, WA, 98195, USA
\end{center}

\section{Additional Simulation study}

We conduct a similar simulation study to section  where the total number of possible covariates is $p=20$. All the other scenarios and coefficients follow the same structure to the simulation study discussed in. 

\begin{table}[!h]
    \centering
\scalebox{1}{
    \begin{tabular}{c|c|cccccccc}
    \toprule
    \multicolumn{2}{c}{$p$} & \multicolumn{8}{c}{20}  \\
    \hline
    \multicolumn{2}{c}{$p(\bgam)$} &
     \multicolumn{8}{c}{Beta-Binomial(1,1)}  \\
  \hline
    \multicolumn{2}{c}{$p_{\bgam_T}$} &
     \multicolumn{2}{c}{0} & \multicolumn{2}{c}{5} & \multicolumn{2}{c}{10} & \multicolumn{2}{c}{20}   \\
  \cmidrule(lr){1-2}\cmidrule(lr){3-4}\cmidrule(lr){5-6} \cmidrule(lr){7-8}\cmidrule(lr){9-10}
     \multicolumn{2}{c}{$r$} & 0 & 0.75 & 0 & 0.75& 0 & 0.75& 0 & 0.75  \\ \midrule
\multirow{ 3}{*}{$\delta=n$}   & LPEP & 99 & 100 & 65 & 17 & 43 & 1 & 91 & 10 \\ 
  & LCE & 99 & 100 & 64 & 14 & 44 & 1 & 34 & 3 \\ 
    & LCL & 99 & 100 & 65 & 15 & 42 & 1 & 40 & 2 \\ 
\midrule
  \multirow{ 3}{*}{$\delta\sim \text{robust}$} & LPEP & 99 & 100 & 62 & 21 & 34 & 4 & 100 & 100 \\ 
  
  & LCE & 99 & 100 & 61 & 23 & 7 & 3 & 100 & 100 \\ 
  & LCL & 99 & 100 & 63 & 22 & 2 & 2 & 100 & 100 \\ \midrule
   \multirow{ 3}{*}{$\delta\sim \text{hyper g/n}$}  & LPEP & 99 & 100 & 65 & 20 & 38 & 8 & 100 & 100 \\ 
  & LCE & 99 & 100 & 57 & 22 & 0 & 1 & 100 & 100 \\ 
    & LCL & 99 & 100 & 57 & 20 & 0 & 0 & 100 & 100 \\ 
\midrule
  & LASSO & 66 & 64 & 0 & 0 & 0 & 0 & 75 & 42 \\ 
  & SCAD & 74 & 67 & 19 & 2 & 4 & 1 & 64 & 34 \\ 
  & MCP & 75 & 66 & 32 & 6 & 13 & 3 & 61 & 36 \\ 
 \bottomrule
    \end{tabular}}
    \caption{Number of times (\textbf{over 100 replications}) that the \textbf{MAP} model coincides with the true model in the logistic regression. }
\end{table}

\begin{figure}[!ht]
    \centering
    \includegraphics[width=0.9\linewidth,height=0.9\textheight]{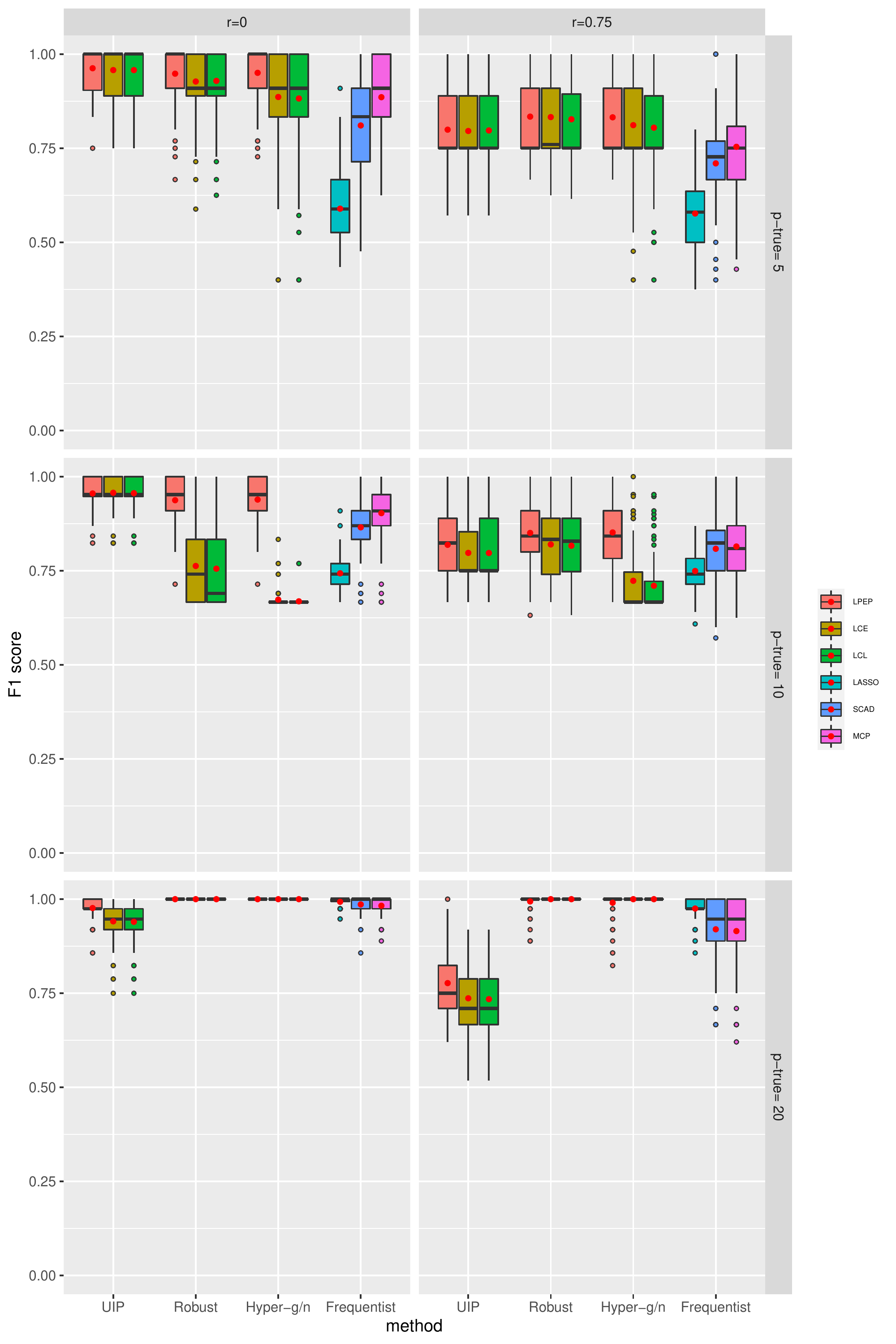}
    \caption{F1 score for the MAP model estimated by various methods and prior combinations for 100 simulated datasets ($n=500,p=20$) under different scenarios of correlation ($r=0$: left; $r=0.75$: right) and true number of non-zero coefficients specified in rows ($p-true=p_{\mathcal{M}_T}$); Red dots represent the average F1 score across 100 simulated datasets.}
    \label{fig:F1score}
\end{figure}

\begin{table}[!h]
    \centering
\scalebox{1}{
    \begin{tabular}{c|c|cccccccc}
    \toprule
    \multicolumn{2}{c}{$p$} & \multicolumn{8}{c}{20}  \\
    \hline
    \multicolumn{2}{c}{$p(\bgam)$} &
     \multicolumn{8}{c}{Beta-Binomial(1,1)}  \\
  \hline
    \multicolumn{2}{c}{$p_{\bgam_T}$} &
     \multicolumn{2}{c}{0} & \multicolumn{2}{c}{5} & \multicolumn{2}{c}{10} & \multicolumn{2}{c}{20}   \\
  \cmidrule(lr){1-2}\cmidrule(lr){3-4}\cmidrule(lr){5-6} \cmidrule(lr){7-8}\cmidrule(lr){9-10}
     \multicolumn{2}{c}{$r$} & 0 & 0.75 & 0 & 0.75& 0 & 0.75& 0 & 0.75  \\ \midrule
\multirow{ 3}{*}{$\delta=n$} & LPEP & 0.53 & \textbf{0.46}* & 10.65 & \textbf{29.57} & \textbf{28.11} & \textbf{59.16} & 54.58 & \textbf{92.20} \\ 
&  LCE & 0.53 & \textbf{0.46}* & 11.28 & 30.27 & 32.18 & 62.77 & 72.08 & 100.45 \\ 
&  LCL & \textbf{0.52}* & \textbf{0.46}* & \textbf{10.31} & 29.71 & 25.22 & 61.38 & \textbf{45.88} & 97.20 \\ 
\midrule
\multirow{ 3}{*}{$\delta\sim \text{robust}$} &  LPEP & 0.56 & \textbf{0.46}* & \textbf{9.59}* & \textbf{25.29} & \textbf{21.48}* & \textbf{44.15}* & \textbf{29.95}* & \textbf{55.92} \\ 
&  LCE & 0.57 & 0.47 & 12.50 & 26.26 & 39.84 & 52.58 & 61.04 & 69.08 \\ 
&  LCL & \textbf{0.52}* & \textbf{0.46}* & 10.78 & 25.79 & 27.99 & 47.38 & 37.15 & 58.29 \\ 
\midrule
\multirow{ 3}{*}{$\delta\sim \text{hyper g/n}$}&  LPEP & 0.68 & 0.69 & \textbf{9.77} & 25.25 & \textbf{22.09} & 45.43 & 30.91 & 57.60 \\ 
&  LCE & 0.67 & 0.61 & 11.40 & 25.19 & 26.64 & 46.26 & 32.64 & 54.10 \\ 
&  LCL & \textbf{0.60} & \textbf{0.49} & 10.73 & \textbf{25.10}* & 24.00 & \textbf{44.65} & \textbf{30.16} & \textbf{53.33}* \\ 
\midrule
&  LASSO & 1.03 & \textbf{1.15} & 16.52 & 30.55 & 28.15 & \textbf{49.72} & \textbf{30.81} & \textbf{58.63} \\ 
&  SCAD & 0.89 & 2.27 & 11.21 & \textbf{30.44} & 26.13 & 59.73 & 45.37 & 75.99 \\ 
&  MCP & \textbf{0.87} & 2.65 & \textbf{10.57} & 32.27 & \textbf{25.69} & 59.97 & 45.05 & 75.43 \\  \bottomrule
    \end{tabular}}
    \caption{1000 times the \textbf{AMSE} for estimated coefficients over 100 replications; \textbf{BOLD} represent group minimum; * represent overall minimum.}
\end{table}

\begin{figure}[ht]
    \centering
    \includegraphics[width=0.9\linewidth,height=0.9\textheight]{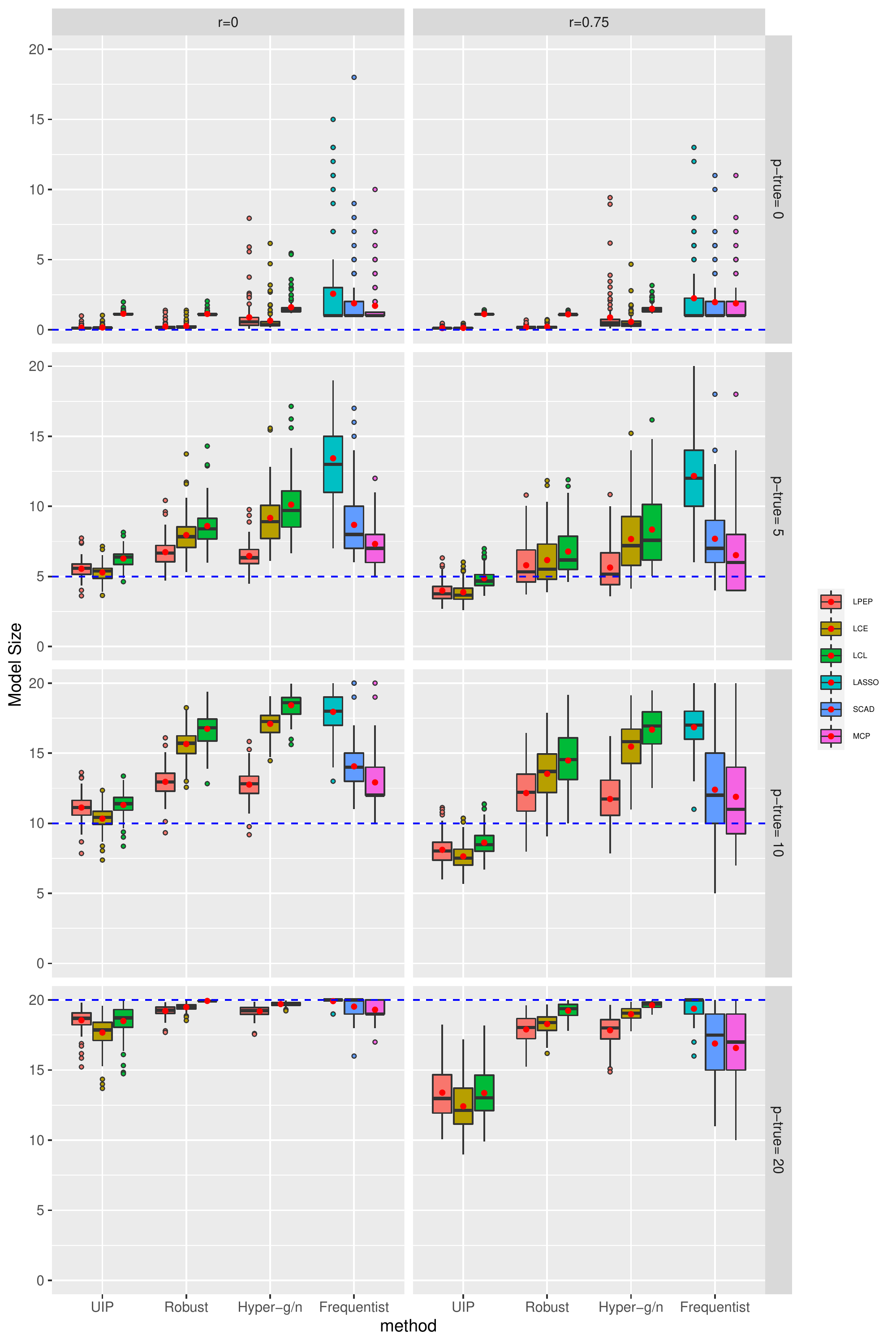}
    \caption{Average size of models selected by various methods and prior combinations for 100 simulated datasets ($n=500,p=20$) under different scenarios of correlation ($r=0$: left; $r=0.75$: right) and true number of non-zero coefficients specified in rows ($p-true=p_{\mathcal{M}_T}$); Dotted blue line indicates the true model size $p_{\mathcal{M}_T}$ and red dots represent the average model size over 100 simulated datasets.}
    \label{fig:modsize}
\end{figure}



\clearpage

\section{\texttt{Endometrial} data set}

The second data set with full separation that we consider is \texttt{endometrial}, which contains data on histology grade and risk factors for 79 cases of endometrial cancer.  This data set was first published in \cite{heinze2002solution}, and is also discussed in \cite{agresti2015foundations}.  In this case, the first variable fully separates the three classes, and the maximum likelihood estimator for the corresponding coefficient is infinite as shown in table \ref{tab:endocoef}.

\begin{table}[!h]
    \centering
    \begin{tabular}{c|c|cccc}
    \toprule
 \multicolumn{2}{c}{} & $\beta_0$ & $\beta_1$& $\beta_2$ & $\beta_3$\\\midrule
 & \multirow{2}{*}{MLE} & 4.30 & 18.19 & -0.04 & -2.90 \\
 & & {\footnotesize(1.43 , 7.95)}  & 
 {\footnotesize(-69.23 , 506.19)} & 
 {\footnotesize(-0.14 , 0.04)} &
 {\footnotesize(-4.79 , -1.44)}\\
 \midrule
 \multirow{8}{*}{$\delta=n$} & \multirow{2}{*}{LPEP} & 3.86& 5.85 & -0.02 & -2.87 \\
 & & {\footnotesize(1.27 , 6.98)}  & {\footnotesize(0.00 , 14.23)} & {\footnotesize(-0.11 , 0.02)} &  {\footnotesize(-4.62 , -1.42)}\\\cmidrule{3-6}
 & \multirow{2}{*}{LCL} & 3.83  & 16.59  & -0.01  & -2.85  \\
 & & {\footnotesize(0.81 , 6.77)} & {\footnotesize(-3317.71 , 3429.10)} & {\footnotesize(-0.11, 0.02)} & {\footnotesize(-4.54 , -1.12)} \\\cmidrule{3-6}
  & \multirow{2}{*}{CRPEP} & 4.14 & 29.72 & -0.02 & -3.04 \\
 & & {\footnotesize (1.53 , 7.40)} & 
 {\footnotesize(1.71 , 25.05)} & 
 {\footnotesize(-0.11 , 0.02)} &  {\footnotesize(-4.81 , -1.53)}\\\cmidrule{3-6}
 & \multirow{2}{*}{DRPEP} & 3.80 & 15.49 & -0.02 & -2.86 \\
 & & {\footnotesize (1.73 , 7.08)} & 
 {\footnotesize(2.47 , 25.96)} & 
 {\footnotesize(-0.13 , 0.03)} &  
 {\footnotesize(-4.51 , -1.62)}\\
\midrule
\multirow{ 4}{*}{$\delta\sim \text{robust}$} & \multirow{2}{*}{LPEP}  & 3.84  & 8.25  & -0.02  & -2.88 \\ 
& & {\footnotesize(1.31 , 7.08)} & {\footnotesize(0.00 , 32.42)} &  {\footnotesize(-0.11 , 0.02)} &  {\footnotesize(-4.75 , -1.41)}\\
\cmidrule{3-6}
 & \multirow{2}{*}{LCL}  & 3.92  & 16.85 & -0.02  & -2.80  \\
 & & {\footnotesize(0.83 , 6.86)} & {\footnotesize(-3303.78 , 3333.49)} & {\footnotesize(-0.12 , 0.03)} &  {\footnotesize(-4.50 , -1.15)}\\
  \midrule
\multirow{8}{*}{$\delta\sim \text{hyper g/n}$} & \multirow{2}{*}{LPEP}  & 3.86   & 6.03   & -0.02 & -2.89  \\ 
& & {\footnotesize(1.25 , 7.13)} & {\footnotesize(0.00 , 19.80)} & {\footnotesize(-0.11 , 0.02)} &  {\footnotesize(-4.67 , -1.40)}\\
\cmidrule{3-6}
 & \multirow{2}{*}{LCL}  & 3.94   & 15.53  & -0.02  & -2.54 \\ 
 & & {\footnotesize(0.84 , 7.02)} & {\footnotesize(-3158.79 , 3255.70)} & {\footnotesize(-0.11 , 0.03)} & {\footnotesize(-4.06 , -0.87)} \\ \cmidrule{3-6}
  & \multirow{2}{*}{CRPEP} & 3.83 & 14.36 & -0.02 & -2.90   \\
 & & {\footnotesize (1.09 , 6.77)} & 
 {\footnotesize(1.92 , 34.42)} & 
 {\footnotesize(-0.11 , 0.02)} &  {\footnotesize(-4.71 , -1.35)}\\\cmidrule{3-6}
 & \multirow{2}{*}{DRPEP} & 4.06 & 9.97 & -0.02 & -3.00\\
 & & {\footnotesize (1.47 , 7.33)} & 
 {\footnotesize(0.00 , 20.56)} & 
 {\footnotesize(-0.12 , 0.01)} &  
 {\footnotesize(-4.82 , -1.45)}\\
  \midrule
 & LASSO & 2.39 & 2.16 & 0.00 & -2.04 \\ 
 & SCAD & 3.02 & 1.84 & 0.00 & -2.44 \\ 
 & MCP & 2.70 & 1.71 & 0.00 & -2.23 \\ \bottomrule
 \end{tabular}
    \caption{Estimated BMA coefficients and 95\% credible intervals for  different Bayesian techniques for \texttt{endometrial} dataset; For frequentist techniques, estimated coefficient is displayed.}
    \label{tab:endocoef}
\end{table}

Similar to the \texttt{urinary} dataset, the regression coefficients for various Bayesian and penalized likelihood methods are shown below along with 95\% credible intervals for Bayesian techniques. For the first coefficient, we again observe that LCL and CR/DRPEP methods produces large estimates with very wide credible intervals under all hyper priors. The can be attributed to the ill-definition of the two priors when the MLE are not finite and data is separable.

We also show the posterior probabilities of all eight models in the model space in Table \ref{tab:endomodel}. The models that are favored by all methods are the full model and model without the second variable. The former is the preferred model by LCL, except when $\delta=n$ while the latter is consistently the preferred model by CRPEP, DRPEP and LPEP.

\begin{table}[!h]
    \centering\scalebox{0.9}{
    \begin{tabular}{c|c|cccccccc}
    \toprule
\multicolumn{2}{c}{} & \multicolumn{8}{c}{$(\gamma_1,\gamma_2,\gamma_3)$}\\\cmidrule{3-10}
 \multicolumn{2}{c}{} & $(0,0,0)$ & $(1,0,0)$ & $(0,1,0)$ &$(1,1,0)$ &$(0,0,1)$ &$(1,0,1)$ &$(0,1,1)$ &$(1,1,1)$  \\\midrule
 \multirow{ 4}{*}{$\delta=n$} & LPEP & 0.00 & 0.00 & 0.00 & 0.00 & 0.02 & \textbf{0.55} & 0.00 & 0.42 \\ 
 & LCL & 0.00 & 0.00 & 0.00 & 0.00 & 0.06 & \textbf{0.60} & 0.01 & 0.33\\ 
  & CRPEP & 0.00 & 0.00 & 0.00 & 0.00 & 0.01 & \textbf{0.52} & 0.00 & 0.47\\
 & DRPEP & 0.00 & 0.00 & 0.00 & 0.00 & 0.00 & \textbf{0.57} & 0.00 & 0.43 \\
\midrule
\multirow{ 2}{*}{$\delta\sim \text{robust}$} & LPEP & 0.00 & 0.00 & 0.00 & 0.00 & 0.03 & \textbf{0.59} & 0.00 & 0.37\\ 
 & LCL & 0.00 & 0.00 & 0.00 & 0.00 & 0.04 & 0.47 & 0.01 & \textbf{0.49} \\ 
  \midrule
\multirow{4}{*}{$\delta\sim \text{hyper g/n}$} & LPEP &  0.00 & 0.00 & 0.00 & 0.00 & 0.03 & \textbf{0.54} & 0.01 & 0.42\\ 
 & LCL & 0.00 & 0.00 & 0.00 & 0.00 & 0.02 & 0.43 & 0.00 & \textbf{0.54} \\ 
  & CRPEP & 0.00 & 0.00 & 0.00 & 0.00 & 0.01 & \textbf{0.62} & 0.00 & 0.38 \\
 & DRPEP & 0.00 & 0.00 & 0.00 & 0.00 & 0.06 & \textbf{0.59} & 0.01 & 0.34\\
  \midrule
 & LASSO & 0.00 & 0.00 & 0.00 & 0.00 & 0.00 & \textbf{1.00} & 0.00 & 0.00 \\ 
 & SCAD & 0.00 & 0.00 & 0.00 & 0.00 & 0.00 & \textbf{1.00} & 0.00 & 0.00 \\ 
 & MCP & 0.00 & 0.00 & 0.00 & 0.00 & 0.00 & \textbf{1.00} & 0.00 & 0.00 \\  \bottomrule
    \end{tabular}}
    \caption{Posterior model probabilities for all the models in the model space for \texttt{endometrial} dataset.}
    \label{tab:endomodel}
\end{table}

\clearpage

\clearpage
\section{\texttt{Pima} data set}

We consider publicly available \texttt{Pima} Indians diabetes dataset \citep{smith1988using}, containing $n=532$ women with complete records on diabetes presence, to illustrate model inference using LPEP prior. This data is publicly available in \texttt{MASS} package. It consists of seven potential covariates to model binary response variable \texttt{type} indicating diabetes presence; namely, number of pregnancies (\texttt{npreg}), plasma glucose concentration (\texttt{glu}), diastolic blood
pressure (\texttt{bp}), triceps skin fold thickness (\texttt{skin}), body mass index (\texttt{bmi}), diabetes pedigree function (\texttt{ped}) and age (\texttt{age}). For Bayesian procedures LPEP, CRPEP and DRPEP, we run MCMC chain for 40000 iterations after a burn-in of 10000 iterations while for LCL, we use full enumeration of models for model selection inference.    

\begin{figure}[!h]
    \centering
    \includegraphics[width=\textwidth]{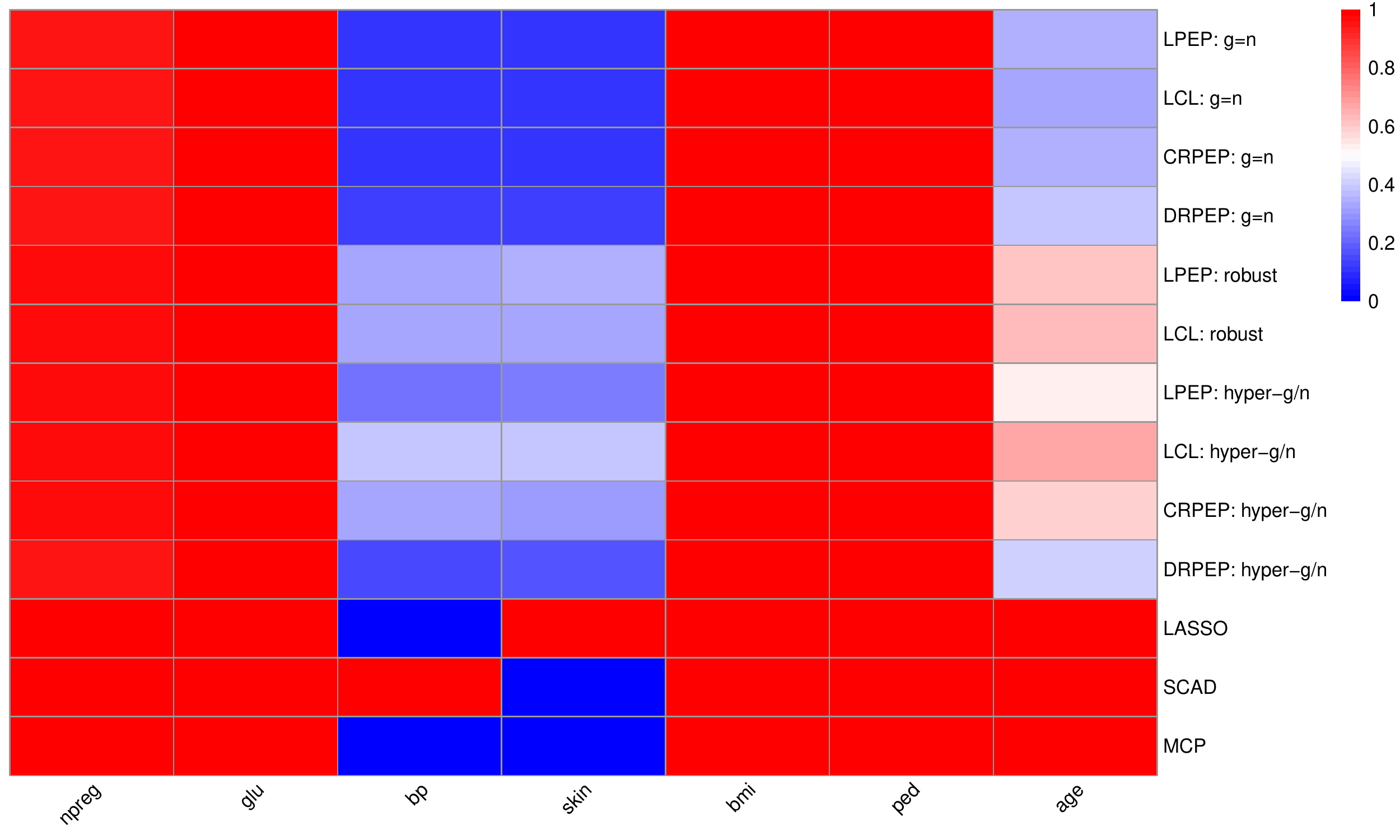}
    \caption{Marginal posterior inclusion probabilities (PIPs) for \texttt{Pima} dataset (Bayesian procedures) and variables included in the model (penalized likelihood methods).}
    \label{fig:pima}
\end{figure}

The marginal posterior inclusion probabilities (PIPs) for each variable are presented in Figure \ref{fig:pima}. Similar to simulation studies, we observe that frequentist techniques select denser model in comparison to Bayesian techniques. Among Bayesian techniques, methods with UIP prior selects more parsimonious model than hyper-g/n and robust counterparts. We also observe that PIP for \texttt{age} variable is not in agreement under the two CRPEP versions.

\clearpage

\section{Details of \texttt{GUSTO-I} dataset}

\begin{table}[!h]
    \centering
    \begin{tabular}{l|l}\toprule
    Variable & Description \\\midrule
    y & Death within 30 days after acute myocardial infarction (Yes = 1, No = 0)\\
    \texttt{SEX} &  Gender (Female=1; Male =0)\\
    \texttt{AGE} &  Age (in years)\\
    \texttt{KILLIP} &  KILLIP class (4 categories)\\
    \texttt{DIA} &  Diabetes (Yes =1; No=0)\\
    \texttt{HYP} &  Hypotension (Yes=1; No=0)\\
    \texttt{HRT} &  Tachycardia (Yes=1; No=0)\\
    \texttt{ANT} &  Anterior infarct location (Yes=1; No=0) \\
    \texttt{PMI} & Previous myocardial infarction (Yes=1, No=0) \\
    \texttt{HEI} & Height (in cm)  \\
    \texttt{WEI} &  Weight (in kg)\\
    \texttt{HTN} &  Hypertension history (Yes=1; No=0)\\
    \texttt{SMK} &  Smoking status (Never/ ex/ current) \\
    \texttt{LIP} &  Hypercholestrolaemia (Yes=1, No=0)\\
    \texttt{PAN} & Previous angina pectoris (Yes=1, No=0) \\
    \texttt{FAM} & Family history of myocardial infarctions (Yes=1, No=0) \\
    \texttt{STE} &  ST elevation on ECG: number of leads (0-11)\\
    \texttt{TTR} &  Time to relief of chest pain more than one hour (Yes=1, No=0)\\\bottomrule

    \end{tabular}
    \caption{Description of variables in \texttt{GUSTO-I} dataset.}
    \label{tab:gusto}
\end{table}

\clearpage
\section{Details of \texttt{HOUSE107} dataset}

\begin{table}[!h]
    \centering
    \scalebox{0.8}{
    \begin{tabular}{l|l}\toprule
    Variable & Description \\\midrule
    y & Legislator voting preference remain unchanged across three vote types (Yes = 1, No = 0)\\
    \texttt{age} &  Age (in years)\\
    \texttt{gender} &  Gender (Female=0; Male =1)\\
    
    \texttt{party.x} & Party of the legislator (0=Democrat; 1= Republican)  \\
    \texttt{numberTerms} & Number of terms served in the House\\
    \texttt{daysServed} & Number of days served \\
    \texttt{black} &  Legislator is African American (Yes=1; No=0)\\
    \texttt{hispanic} & Legislator is Hispanic (Yes=1; No=0) \\
    \texttt{numCosp} & Number of bills an MC cosponsored in a term  \\
    \texttt{abroadPrcnt} & Percent of the district that lived abroad in the census used for this term  \\
    \texttt{recentArrivalPrcnt} & Percent of the district that recently moved into the district from another county
or state \\
    \texttt{prcntForeignBorn} & Percentage of the population in the district that are foreign born\\
    \texttt{prcntExAliens} & Percentage of the foreign born population in the district that became a citizen  \\
    \texttt{medianIncome} & Median income of district represented by the legislator\\
    \texttt{prcntUnemp} & Unemployment rate in the district represented by the legislator\\
    \texttt{prcntNotEmploy} &  Percentage of adults not employed in the district represented by the legislator\\
    \texttt{prcntBA} & Percentage of the adult  population in district with bachelor´s degree\\
    \texttt{prcntHS} & Percentage of the adult  population in district with a high school degree \\
    \texttt{prcntAsian} & Percentage of Asian population in district represented by legislator\\
    \texttt{prcntBlack} & Percentage of African American population in district represented by legislator\\ 
    \texttt{prcntHisp} & Percentage of Hispanic population in district represented by legislator\\
    \texttt{prcntMulti} & Percentage of population with multiple ethnicity in district represented by legislator\\
    \texttt{prcntWhite} & Percentage of White population in district represented by legislator, excluding hispanics \\
    \texttt{prcntWhiteAll} & Percentage of White population in district represented by legislator, including hispanics \\
    \texttt{prcntOld} & Percentage of individual over 65 years of age in district represented by legislator\\
    \texttt{medianAge} & Median age of  district represented by legislator\\
    \texttt{gini} & Gini Index of the district represented by the legislator\\
    \bottomrule

    \end{tabular}}
    \caption{Description of variables in \texttt{HOUSE107} dataset.}
    \label{tab:house107}
\end{table}

\bibliographystyle{apalike}

\bibliography{LPEP}